\documentclass[%
 reprint,
 amsmath,amssymb,
 aps,
]{revtex4-2}

\usepackage{graphicx}
\usepackage{subcaption}
\usepackage[T1]{fontenc}
\usepackage{textcomp}
\usepackage[utf8x]{inputenc}
\usepackage{amsmath}
\usepackage{amssymb}
\usepackage{dcolumn}
\usepackage{bm}

\begin{document}

\preprint{APS/123-QED}

\title{The negative dependence of evacuation time on group size under a binding mechanism}

\author{Tianyi Wang}
  \email{tianywang18@fudan.edu.cn}
\affiliation{
Department of Physics, Fudan University
}

\author{Yu Chen}%
 \email{chen@edu.k.u-tokyo.ac.jp}
\affiliation{
Department of Human and Engineered Environmental Studies, Graduate School of Frontier Science, The University of Tokyo, Kashiwa, Japan
}%

\date{\today}

\begin{abstract}
This paper initiates the analysis of the relation between evacuation time and group size by applying an extended floor-field cellular automaton model. Multi-speed-agents, a group structure containing leaders and followers, and a local-density-dependent dynamic field are implemented all together in the model. Most importantly, a complete binding mechanism which includes leaders’ waiting for followers is brought up for the first time. A counterintuitive negative relation between evacuation time and group size is discovered in one-group-type simulations.  An entropy-like quantity, namely the mixing index, is constructed to analyze the cause of that relation. It is found that under the binding mechanism, the higher degree of group mixing, the longer the evacuation time will be. Moreover, through a constant scale transformation, it is shown that the mixing index can be a key indicator that contains useful information about the evacuation system.

\end{abstract}

\maketitle

\section{\label{sec:level1}Introduction\protect}

Pedestrian and evacuation dynamics (PED) is a topic that stems from practical situations in human society as well as from non-equilibrium physics \cite{kirchner_simulation_2002,luo_update_2018}, which is attracting more and more interests from sociologists,  computer scientists and physicists.  Various models have been applied to simulation studies of PED\cite{li_review_2019}, including the social force model \cite{helbing_social_1995,mehran_abnormal_2009,anvari_modelling_2015,hou_social_2014}, the centrifugal force model \cite{yu_centrifugal_2005,chraibi_generalized_2010,chraibi_force-based_2011,chraibi_modeling_2012}, the Reciprocal Velocity Obstacles model \cite{van_den_berg_reciprocal_2008} and the cellular automaton (CA) model \cite{fu_floor_2015,lu_study_2017,pereira_emergency_2017,lu_modeling_2014}, etc., among which the CA model stands out not only for its computational efficiency but also for  its higher extendibility \cite{wei-guo_evacuation_2006}. 

Group structure has been an important issue in PED \cite{james_distribution_1953,fang_leaderfollower_2016,muller_study_2014} yet not fully understood through present studies. In most cases of evacuation, instead of acting like independent individuals, people tend to make their efforts to evacuate with their family, close friends, or even acquaintances at the site. Such action would lead to the formation of a number of small groups on the evacuation spot. Regarding the evacuating behavior, agents belonging to the same group would be more likely to stay together, a conceivable phenomenon that we categorize  as the effects of “binding ” in this article. Lili Lu et al. employed a floor-field-based CA model including the group structure by exploiting the extendibility of the CA model\cite{lu_study_2017,lu_modeling_2014}. Through both numerical and field experiments in their study, the authors found that basically evacuation time has no relation with group size. However, we want to point out that this observation needs to be investigated carefully since only one part of the binding effects, that is, group members’ following their leaders, is considered in Lili Lu’s model.  Another part of the binding effects, namely leaders’ waiting for the followers, should also be taken into account.

In this study, we aim to conduct a detailed study on the binding effects in a group-structured PED with an extended floor-field CA model. The construction of floor-field and efficiency function for agents to decide their motion is based on the work of Lili Lu et al \cite{lu_study_2017}, though we also made a series of extensions including the implementation of diverse walking abilities for agents \cite{fu_floor_2015} and the adoption of a local-density-dependent dynamic floor field. The latter extension can be justified by considering the necessary condition for the occurrence of herding behavior in the evacuation. Most importantly, while the mechanism of following is kept unchanged, we added a waiting mechanism among the group members such that group leaders would not move forward until they could find their followers within a predetermined distance. 

Based on qualitative and quantitative analyses, we find, from a large number of simulations of the one-group-type evacuation, an unexpected negative dependence of evacuation time on group size. We show that such a negative dependence actually originates from the binding mechanism. As the binding weakens, the negative dependence gets weaker. To explain the deviation from our intuition (that the evacuation time of larger groups should be longer than that of small groups under the binding mechanism), we proposed an entropy-like quantity, namely the “mixing index” , by applying a transformation of representation \cite{nazarov_advanced_2013} from agents to grids. We reveal that the mixing index will decrease as the group size gets larger, which can serve as the cause of the negative dependence of evacuation time on group size.

The body of the paper is organized as follows. Sec.II is devoted to the construction of the extended floor-field based CA model with the complete binding mechanism. The validation of our model is also shown in the same section. Sec.III exhibits the simulation results from which the relation between evacuation time and group size is found. The definition of mixing index is introduced in Sec.IV, with which the cause of negative dependence of evacuation time on group size is explained. Finally,  the conclusion of the whole study is given in Sec.V.

\section{\label{sec:level1}Model\protect}

\subsection{\label{sec:level2}General description}

Basic components of our model include the evacuating agents and floors of the evacuation area. The floor is discretized into  $(L+2)(L+2)$ grids of an equal size. The evacuating agents can only move in the central $L \times L$ area. Boundaries of the whole area are denoted as $(0, k), (L+1,k), (k, 0), (k, L+1)$ with $k=0,1,...,L,L+1$. Two exits, each of which has a width of l grids, are symmetrically positioned on the right boundary with the distance of $l$ grids from upper (lower) boundary, see Fig. 1. 

\begin{figure}[htbp]
\includegraphics[width=8cm]{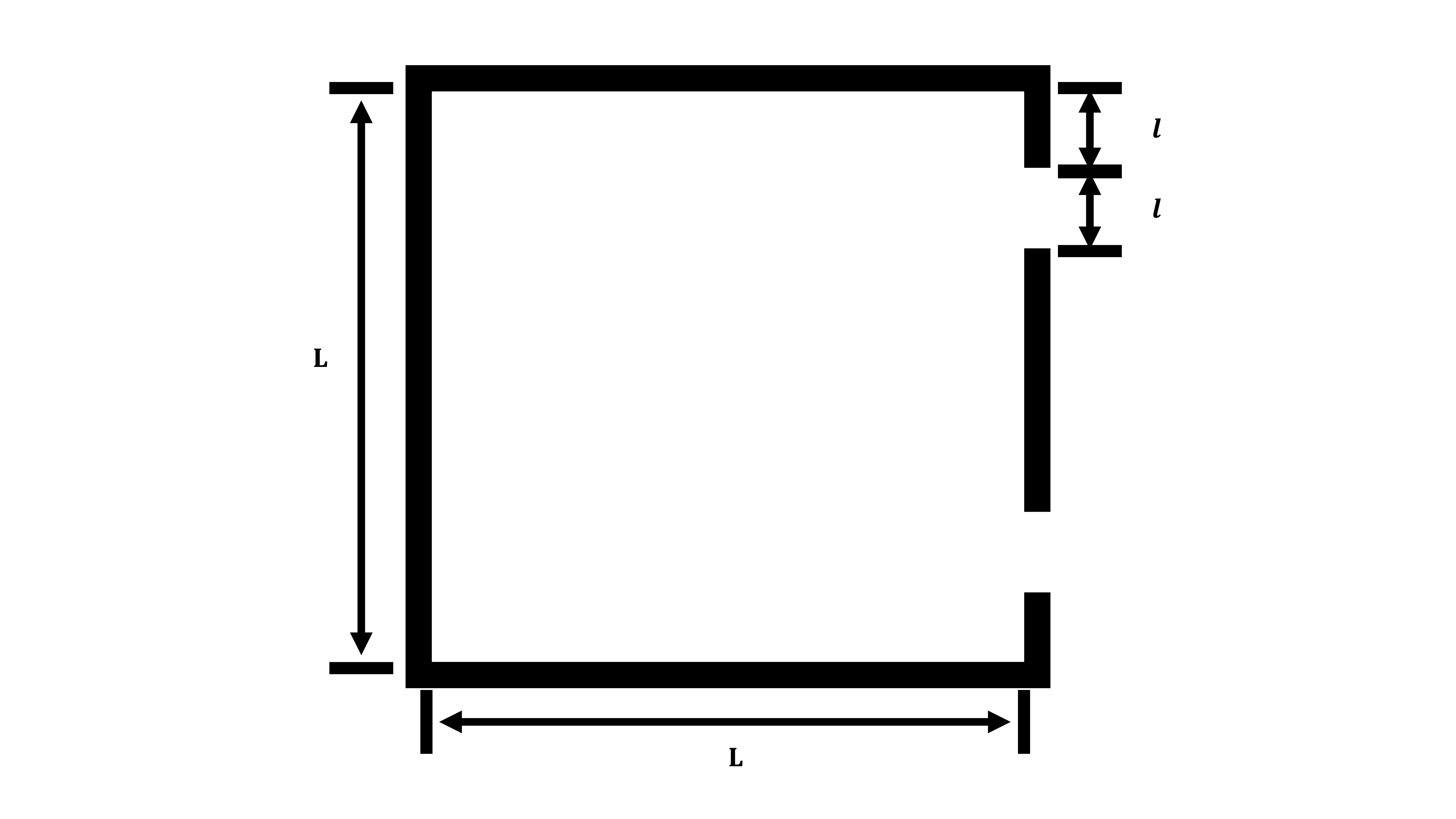}
\centering
\caption{\label{fig:epsart}Configuration of evacuation area.}
\end{figure}

There exist three types of evacuating agents, assigned with different moving speeds: $v = 1$, $2$, $3$ grids per time-step. Agents with multiple moving speeds (multi-speed-agents hereafter) stand and move on the floor, with their locations labeled by the coordinates of the grids which they are occupying. Note that every agent can occupy only one grid and each grid can accommodate only one agent at a given moment. Hence the existence of an agent on a given grid can be denoted as a binary variable $n^v_{i, j}(t)$, where $i$,$j$ are integers from $1$ to $L$ stand for the grid coordinates (grids on boundaries are forbidden to agents other than gate area), $t$ for time, and $v$ labels the speed. Since a grid can contain at most one agent, $n^1_{i, j}$, $n^2_{i, j}$ and $n^3_{i, j}$ are exclusive to each other. 
To describe the motions of multi-speed-agents, we divide a given time step into three pseudo sub-steps. All the agents, whenever possible, will move in the first pseudo sub-step. Agents with larger speeds may move in the second or the third pseudo sub-step. The direction and the goal of an agent’s motion are determined by an efficiency function at each time-step. The moving process continues until all agents have left the area. The evacuation time of an individual agent is counted as the number of time steps he or she takes to exit, hence the average evacuation time $\bar{T}$ is the evacuation time averaged over all agents. On the other hand, the total evacuation time $T_{tot}$ is defined as the total time steps elapsed until the last evacuating agent left the area through either one of the exits. 

\subsection{Static and dynamic floor-fields}
By analogy with the electric and magnetic fields in physical space, it is assumed that there are floor-fields that will affect the movement of agents through the working of the efficiency function. We assume that the floor-fields have two components: the static floor-field and the dynamic floor-field \cite{lu_study_2017}. 

The strength of static floor-field on grid $(i, j)$ is denoted as $S_{i, j}$, which is determined by the distance from the grid to the nearest exit. The purpose for the introduction of such a static field is to model the behavior of a perfectly rational and fully informed evacuating agent, who takes the direct distance to the exit as the only consideration in making decisions during the evacuation. To be more specific, the static floor field is set by following three steps \cite{lu_study_2017}:
(1) Select a grid and evaluate the distances $d_i$ ($i=1, ..., n$) from the current grid to the two exits through all the n possible direct paths, in reference with Fig. 2. 
(2) Select the shortest path dmin from {$d_{1}$ , ..., $d_{n}$ }, and set the reciprocal of this distance as the static floor-field strength of the grid, i.e. $S_{i, j} = \frac{1}{d_{min}}$.
(3) Repeat (1), (2) for all the grids in the evacuation area.

\begin{figure}[htbp]
\includegraphics[width=4cm]{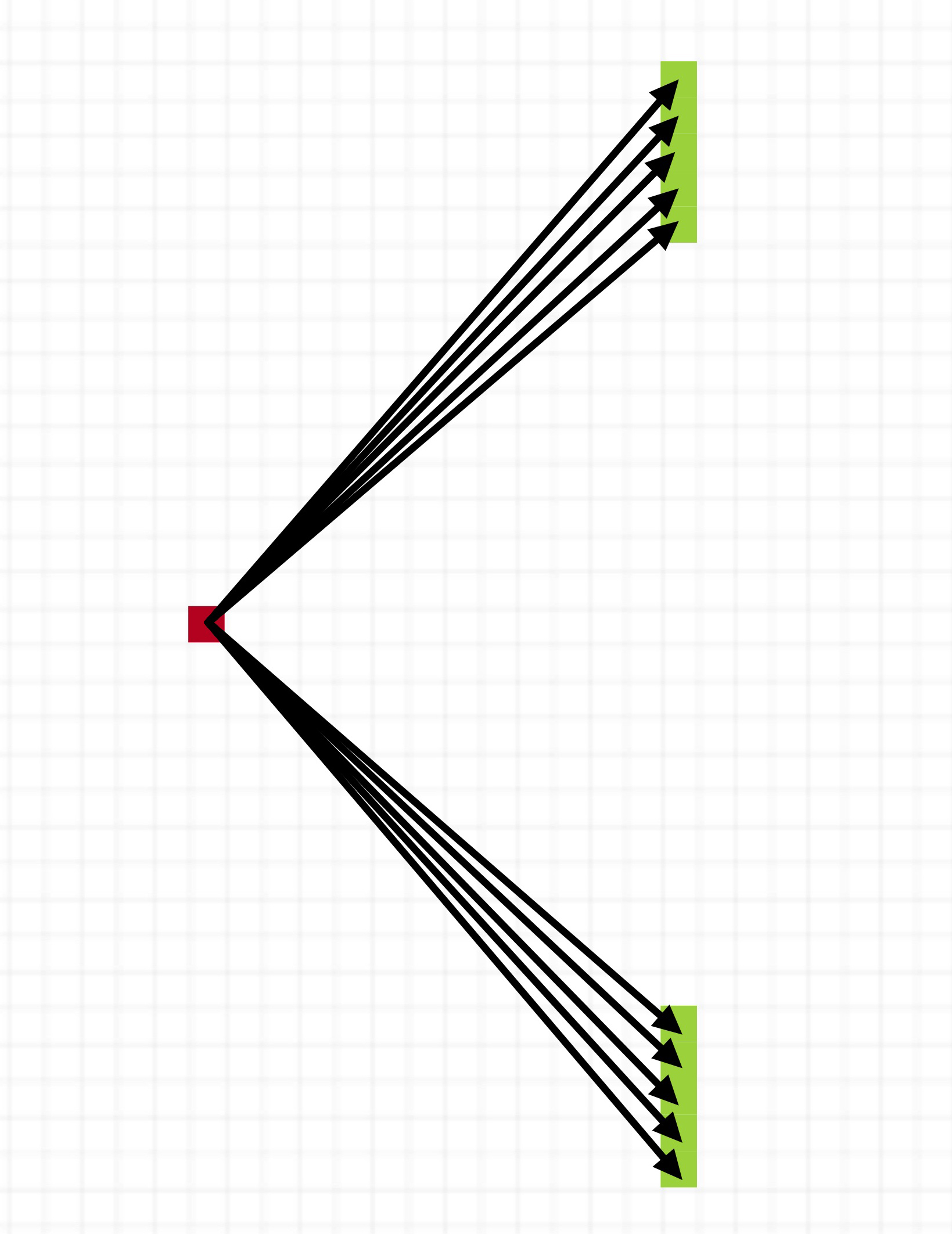}
\centering
\caption{\label{fig:epsart}Distance evaluation of evacuation paths (from top to bottom are labelled sequentially from 1 to 10).}
\end{figure}

The strength of dynamic floor-field on grid $(i, j)$ is denoted as $D_{i, j}$. In contrast with the static floor-field, dynamic floor-field varies with time, which may instantly trigger the herding of agents during the evacuation process. Dynamic floor-field is initialized with a ground value of zero for all the grids, and updated subsequently through the evacuation process by following the scheme below\cite{lu_study_2017}:
(1) For a given empty grid, its dynamic floor-field strength will increase with a unit value if it is occupied by an agent at the following time-step, i.e., $D_{i, j}(t) = D_{i, j}(t-1)+\delta(n_{i,j}(t))\delta(1-n_{i,j}^{v}(t-1))$, where $\delta(x)$ represents a delta function.
(2) For a given occupied grid, its dynamic floor-field strength will decay over a time step with a ratio $0<\alpha<1$, i.e., $D_{i,j}(t)=\alpha D_{i,j}(t-1)$.
(3) For a randomly chosen grid, a portion $0<\theta<1$ of its dynamic floor-field strength will diffuse into grids locating on its Von Neumann neighborhood, i.e., $D_{i\pm1,j\pm1}(t)=D_{i\pm1,j\pm1}(t-1)+\theta D_{i,j}(t-1)$.

\subsection{\label{sec:citeref}Group structure}

\subsubsection{Basic group setting}
Groups of different sizes are differentiated as the 2-agent groups, 3-agent groups, 4-agent groups and 5-agent groups in this study. Only one leader exists in a group, and the rest agents in the group are called followers. Leaders and followers have different types of efficiency functions. For leaders, the efficiency function has the same form as that of the non-grouped individuals. Followers are defined to have more limited rationality and they are less influenced by the static floor-field so that they mainly follow the leaders’ motion under the basic configuration. Lastly, leaders and followers are mutually constrained by the binding mechanism which will be described in detail in Sec.II.E.

\subsubsection{Diverse mobility}
We couple the agents’ diverse mobility to the group structure in order to see its role played in the evacuation dynamics. We approximate the population of agents with different speeds as $N^3:N^2:N^1=2:3:5$, with $N^v=\sum_{i,j}n^v_{i,j}$. Considering that agents equipped with a higher mobility are relatively more capable of taking care of other members in an urgent situation, we further assume that group leaders are always those with the highest walking ability among all the group members. Fig.3 shows an example of distributing 40 multi-speed-agents into 10 groups, each of which consists of 4 evacuating agents.

\begin{figure}[htbp]
\includegraphics[width=8cm]{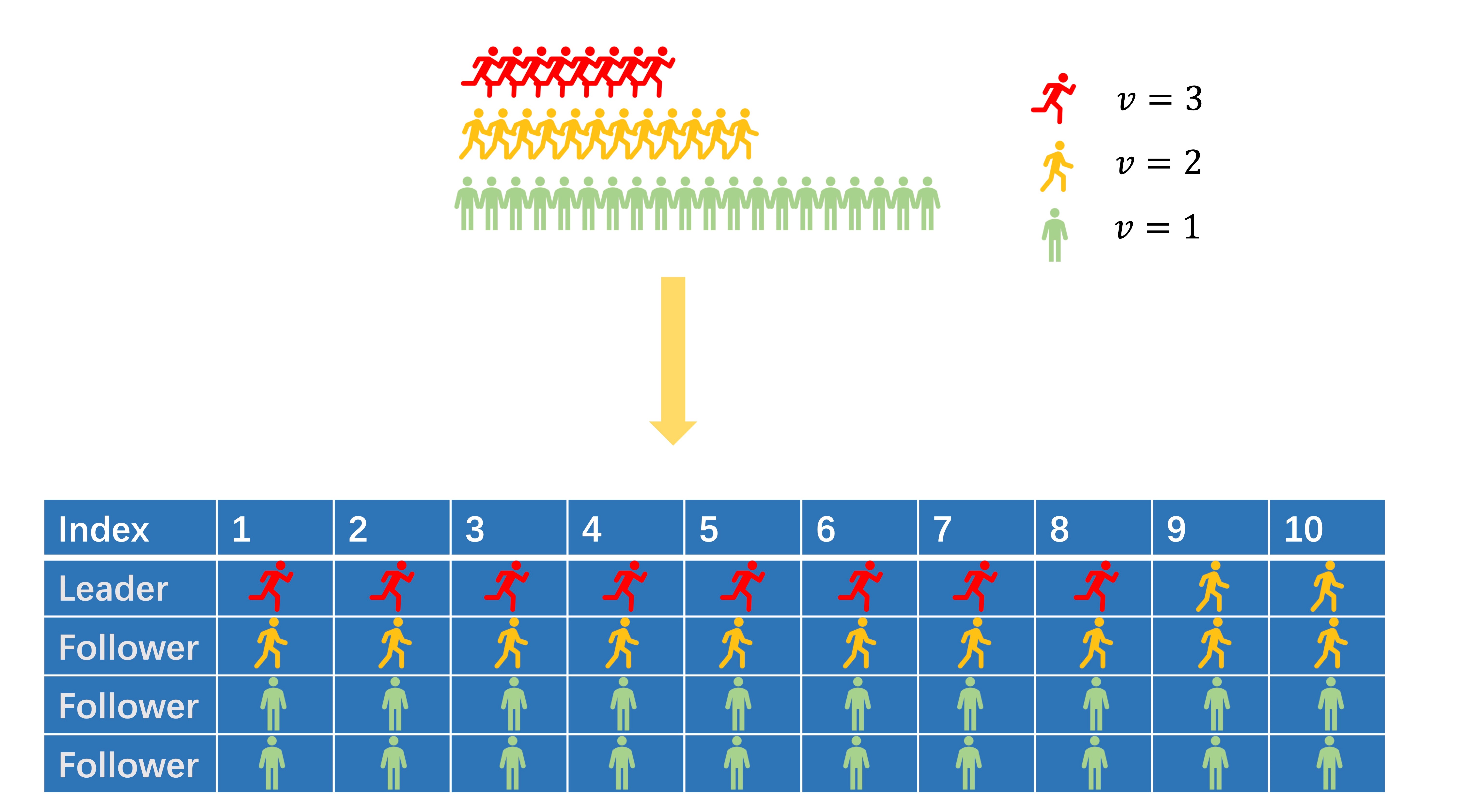}
\centering
\caption{\label{fig:epsart}An example of distributing 40 multi-speed-agents into ten 4-agent groups.}
\end{figure}

\subsection{\label{sec:citeref}Decision making}

\subsubsection{Efficiency function}
At each time-step, agents shall decide which grid to move into with the use of efficiency functions. Inspired by the “no back step” assumption made by Weng et al in their study of pedestrian flow \cite{weng_cellular_2006}, candidates of the goal grids for an agent to locate are always in or at the front of its current grid, along the direction of evacuation. For an agent staying in grid (i, j) (the grid in blue in Fig. 3), the three grids in the Von Neumann neighborhood $\{i',j'\} =\{(i+1,j), (i,j+1), (i, j-1)\}$  (the grids in orange in Fig. 3) are the possible goals. Therefore, the agent shall calculate the efficiency function $F_{i'j'}$  for all the neighboring grids.

\begin{figure}[htbp]
\includegraphics[width=4cm]{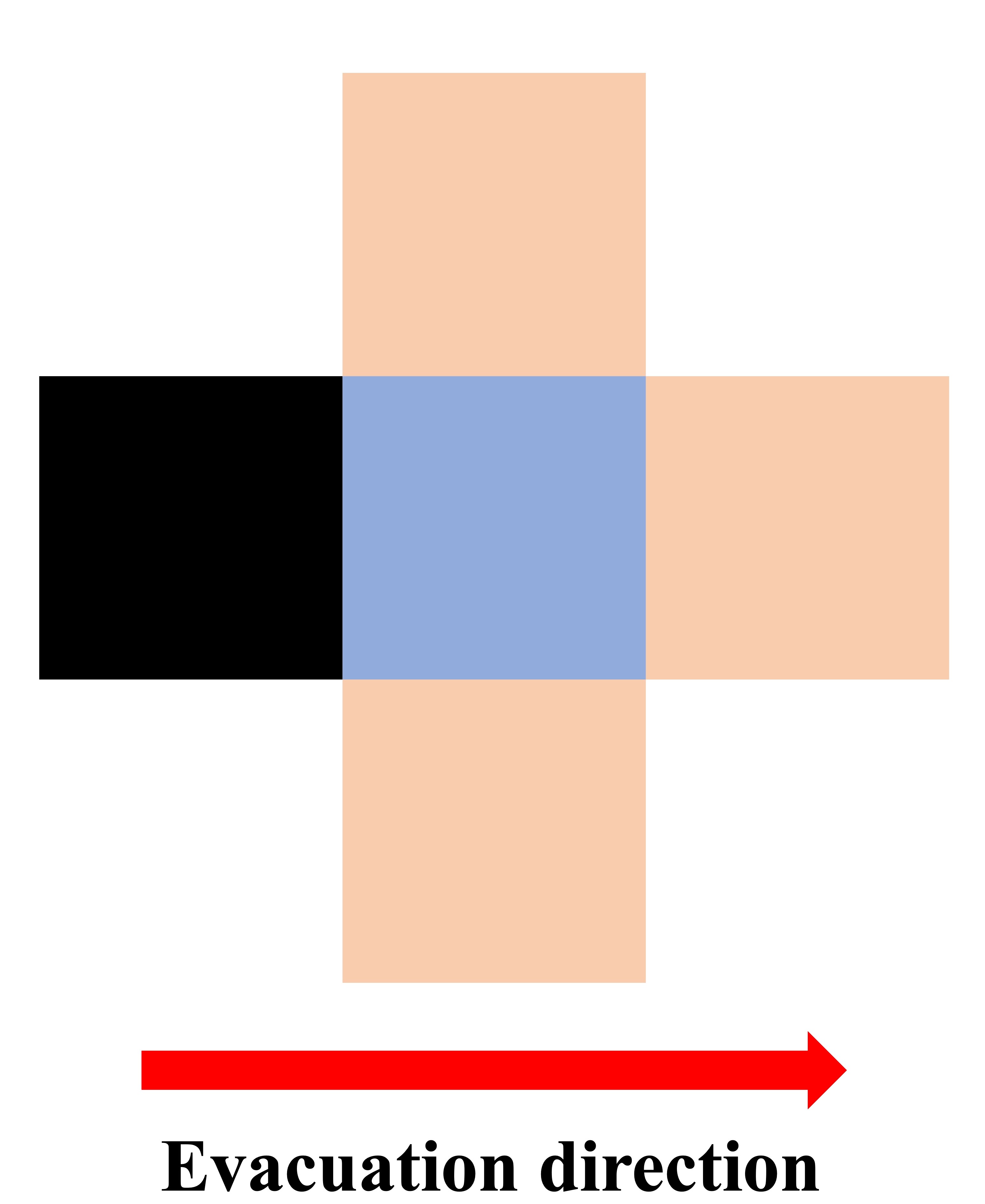}
\centering
\caption{\label{fig:epsart}Candidates of target orange grids for an evacuation agent staying in the central blue grid. Under “no-back-step” assumption, entrance into the black grid is forbidden for the agent.}
\end{figure}

For the leaders and the individual agents, the efficiency function is formulated as the following \cite{lu_study_2017},
\begin{equation}
F_{i'j'} = C exp(k_{S} S_{i' j'}+k_{D} D_{i' j'})(1-\eta_{i' j'})
\end{equation}
For the followers, the efficiency function reads as follows,
\begin{equation}
F_{i' j'} = C exp(k_{S} S_{i' j'}-k_{d}d_{i' j'}+k_{p}\zeta_{i'-i,j'-j})(1-\eta_{i' j'})
\end{equation}
In Eqs.(1) and (2) above, constant C stands for the normalization factor, and a set of positive parameters ${k_S ,k_S' ,k_D ,k_d ,k_p}$are defined as the sensitivity coefficients. Explanations as well as modifications to these parameters are provided as follows. Firstly, the influence of static floor-field is weighted by $k_S$ for leaders and  $k_S'$ for followers respectively. Next, the distance from grid $(i', j')$ to the grid in which the group leader stays is denoted as $d_{i'j'}$ for the followers. The negative $k_{d}d_{i'j'}$ term in the exponential function results in a higher efficiency for grids closer to the leaders, helping the followers get closer to their leaders. Thirdly, the formulation of sensitivity to the dynamic floor-field is modified in our model as  $k_D=k_{D0}H(N(r)-N_0)$, where $k_{D0}$ stands for a constant and $H(x)$ is the Heaviside function. $N(r)$ represent the number of agents within the observation range of radius $r$, and $N_0$ is a threshold. The reason for such a formulation can be explained from two aspects: (1) If the local population surpassed the threshold in the observation range, the evacuating agent (even if he or she can be a group leader) tend to be uncertain about their surroundings so that they would be more likely to follow other agents; (2) This formulation can give a remedy to the unnatural “locking” problem caused by the inter-attractive motion among agents through the influence of dynamic floor-field. In the later stage of evacuation simulation using the Lily Lu’s model with certain parameters,  one can often observe that the remaining few agents will cluster together in distant grids to the exits and perform a back and forth motion instead of moving towards the exits. The reason behind this is that the static floor-field is locally weaker,  hence the dynamic floor-field dominates. In practice, we set $r=4, N_0=2$, hence the overall evacuation dynamics would not be affected by such a modification.  

The remaining two binary variables in Eqs.(1) and (2)  are defined respectively as a flag for the occupation of a grid $\eta_{ij}=\sum_{v}n^v_{i,j}$, and a flag for the alignment of moving direction of leaders and followers $\zeta_{i'-i,j'-j} = \delta(\hat{v}_l-\hat{v}_f(i'-i,j'-j))$, where $v_l$  labels the direction of the group leader’s current velocity, and $v_f(i'-i,j'-j)$ denotes the direction of the follower’s velocity.

In boundary grids, efficiency functions are set as zero except for those on the exits. In the exit grids, efficiency  functions are set to be infinite. With such a disposal of boundary conditions, agents are not able to move out of the evacuation area unless they get evacuated from one of the exits.

\subsubsection{Decision of movement}
By completing the calculation of efficiency functions, the evacuating agent can make his or her decisions by choosing a goal grid with the largest efficiency function for the next movement. 
\begin{equation}
(i, j)_{t'+1}\equiv E_0((i, j)_{t'})=arg max_{i',j'} F_{i'j'}
\end{equation}
In Eqs.(3), $t'$ is the index of the current pseudo time step, and $E_0$ as the error-free position decision function based on the evaluation of efficiency function. With the consideration of agents’ irrationality, a panic-induced error in the decision making process can be modeled through the following two steps.
At each time-step, after an agent has decided which grid to move into, a random number  uniformly distributed in $[0, 1]$ is generated.
Decision of the agent will be adjusted with the following formula
\begin{equation}
(i, j)_{t'+1}\equiv E_1((i, j)_{t'})
\end{equation}

\begin{align}
E_1((i, j)_{t'}) &=
 H(\theta_{err}-\theta)\cdot (i',j')_{Rand} \nonumber \\
 &+ H(\theta-\theta_{err})\cdot E_0((i,j)_{T'})
\end{align}

in which $E_1$ is defined as the error-included decision function, $\theta_{err}$ is the criteria of making mistake, and $(i',j')_{Rand}$ is a goal grid randomly chosen from all the unoccupied grids in $(i-1,j),(i+1,j), (i,j+1), (i, j-1)$. 
The scheme for decision making above can be easily generalized to multi-speed-agents, namely 
\begin{equation}
(i, j)_{t+1}=E_1^{(v)}((i, j)_{t'}).
\end{equation}
In Eqs.(6), $ E_1^{(v)}((i, j)_{T'})=E_1\circ...\circ E_1 ((i,j)_{T'})$, meaning that the application of function $E_1$ is repeated $v$ times for agents with moving speed $v$.

\subsubsection{Binding mechanism}
As is described in Sec.I, we define the empirically observed tendency, that agents of the same group are more likely to cluster together, as the binding mechanism in this model. There are two core effects for the binding, which state that followers always try to follow the leaders of the same groups and leaders will wait for the followers if they are too far behind. 
The following mechanism is implemented via efficiency function $F_{i'j'}$ for the followers, in reference to Eqs.(2). Following is enabled by $k_d$ term related to the relative distance with the leader and the $k_p$ term related to the direction of the leader’s motion in the efficiency function. Followers are more likely to move into grids that are closer to the leaders or to move in directions aligned with those of the leaders. 
The waiting mechanism is implemented for the leaders in the decision of moving directions. In particular, the direction for the next movement of agents are defined by an error-included decision function,
\begin{equation}
(i, j)_{T'+1}\equiv E ((i, j)_{T'}).
\end{equation}
For leaders,
\begin{equation}
E ((i, j)_{T'}) = H(d_0-d_f)\cdot E_1 ((i, j)_{T'}) + H(d_f-d_0)(i, j)_{T'}.
\end{equation}
with
\begin{equation}
d_f=max \{d_k\}.
\end{equation}

For followers,
\begin{equation}
E ((i, j)_{T'}) = E_1 ((i, j)_{T'}).
\end{equation}
In Eqs.(8) and (9), $d_0$ is the critical waiting distance, $\{d_k\}$ records distances from the group leader to each follower for a given group. The characteristic of decision function $E$ is whenever the longest distance from the leader to a follower surpasses $d_0$ , the leader should stay put. 
Note that in the previous study of Lili Lu et al, only a part of the binding effects, namely the following process, is implemented. Hence we may not find the clustering of the group members in the first place under the multi-speed-agents extension. Nevertheless, in our point of view, the interaction between the leader and follower is of fundamental importance, and we expect the clustered structure would have non-trivial influence on the whole PED dynamics. 

\subsection{Model Validation}

\subsubsection{Simulation procedure and initial conditions}
The model used in the current study has been extended in the following aspects: (1) the realization of the full binding effects between group leaders and followers; (2) the inclusion of multi-speed-agents; as well as (3) the remedy of the “locking” problem. Therefore, we need to validate the updated model by showing that the overall PED has not been altered, so that we can provide a solid foundation for the specific study on the influence coming from the group structure on the evacuation. 
The basic procedures in the simulation code are as follows:\\
1.Initialize the population of evacuating agents, the static and dynamic floor-fields.\\
2.In each time-step:\\
(1)Calculate efficiency functions attached to each agent by sweeping the grids from right to left.\\
(2)Make decisions on movements for all the agents and update their positions.\\
(3)Repeat (1) and (2) for the whole three pseudo sub-steps.\\
3.Update the dynamic floor-field and the number of the remaining agents.\\
4.Repeat 1, 2, 3, till there are no agents left in the evacuation area.\\
To be more specific about the initialization of population, locations of the agents in the same group are set to stick together\cite{lu_study_2017}. From each agent’s view, there must be another agent from the same group in the Miller neighborhood. The group leader is randomly chosen. 
For simulations shown in the following sections, the population is set as $N(t=0)=480$ agents and the configuration of grids as $L=40$ and $l=5$. Statistics are obtained by repeating each simulation 100 times and taking the average of the counted numbers.

\subsubsection{Condition I: group structure without  multi-speed-agents}
Lili Lu et al. found that there is no evident correlation between group size and evacuation time\cite{lu_study_2017}. This result can be reproduced when we only take the following process in the binding mechanism into account and exclude the diversity in the moving speeds. We conduct simulations for the evacuation of individuals, 2-agent groups and 3-agent groups, counting the number of agents (indicated by N(t)) left in the room as time evolved. Our results agree with Lili Lu’s finding with parameters shown in row I of Table I. 

In the results shown in Fig. 5, obviously, the total evacuation times for the 2-agent groups and the 3-agent groups are close, both of which are longer than the total evacuation time of individuals.  The corresponding results can be found in the study of Lili Lu et al (see Fig. 8 in \cite{lu_study_2017}). Although there are differences attributing to the intrinsic different setting of models, such as the way movement is decided and the different settings of sensitivities to the dynamic floor-field, etc., the weak correlation between group size and evacuation time has been reproduced.

\begin{table}[h!]
\centering
 \begin{tabular}{||c p{1cm}<{\centering} p{1cm}<{\centering}||} 
 \hline
 Parameters &  I &  II \\ [1ex] 
 \hline
 $k_{S}$ & 8 & 8 \\ 
 $k_{S}'$ & 6 & -  \\
 $k_{D_{0}}$ & 1 & 6 \\
 $k_{p}$ & 1 & - \\
 $k_{d}$ & 3 & - \\
 $\alpha $ & 0.2 & 0.2 \\
 $\delta$ & 0.1 & 0.1 \\
 $d_{l}$ & 4 & 4 \\
 $N_{l}$ & 2 & 2 \\
 $\theta_{err}$ & 0.2 & 0.2 \\ [1ex] 
 \hline
 \end{tabular}
 \caption{Parameter settings in \\ validation condition I and II.}
 \label{table:1}
\end{table}
\begin{figure}[htbp]
\includegraphics[width=8cm]{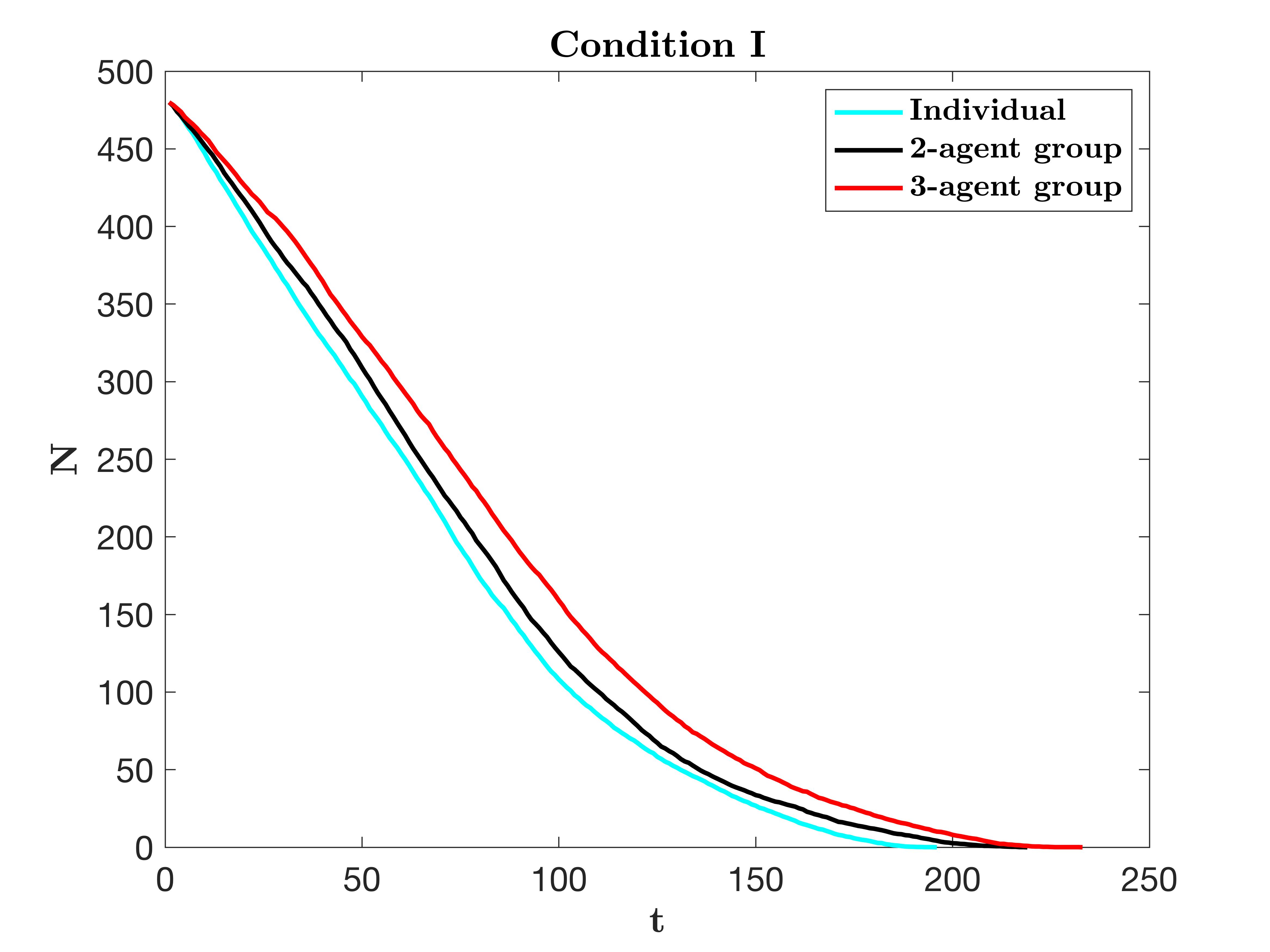}
\centering
\caption{\label{fig:epsart}Time evolution of the number of agents remained in the room
 in condition I.}
\end{figure}

\subsubsection{Condition II:  multi-speed-agents without group structure}
The second validation concerns with the evacuation of individuals with different walking abilities. It is shown in the work of Zhijian Fu et al that the concavity of the evacuation evolution curves differs for agents of different walking speeds. Specifically, the convex curves for  low-speed agents will change into concave curves for the high-speed agents\cite{lu_study_2017}.
We conduct similar evacuation simulations including agents who are uniformly assigned with speed $v$ of 1 to 5 grids per time-step. Parameters are listed in row II of Table I and simulation results are shown in Fig. 6. In a statistical sense, our results agree with that of Fu’s study (see Fig.2 (b) in \cite{lu_study_2017}). The differences between the two models may be caused by the facts that Fu et al use Moore instead of Von Neumann neighborhood in the decision of the next movement, and that they implement the movement of multi-speed-agents with floor-fields instead of the pseudo sub-step algorithm. 
\begin{figure}[htbp]
\includegraphics[width=8cm]{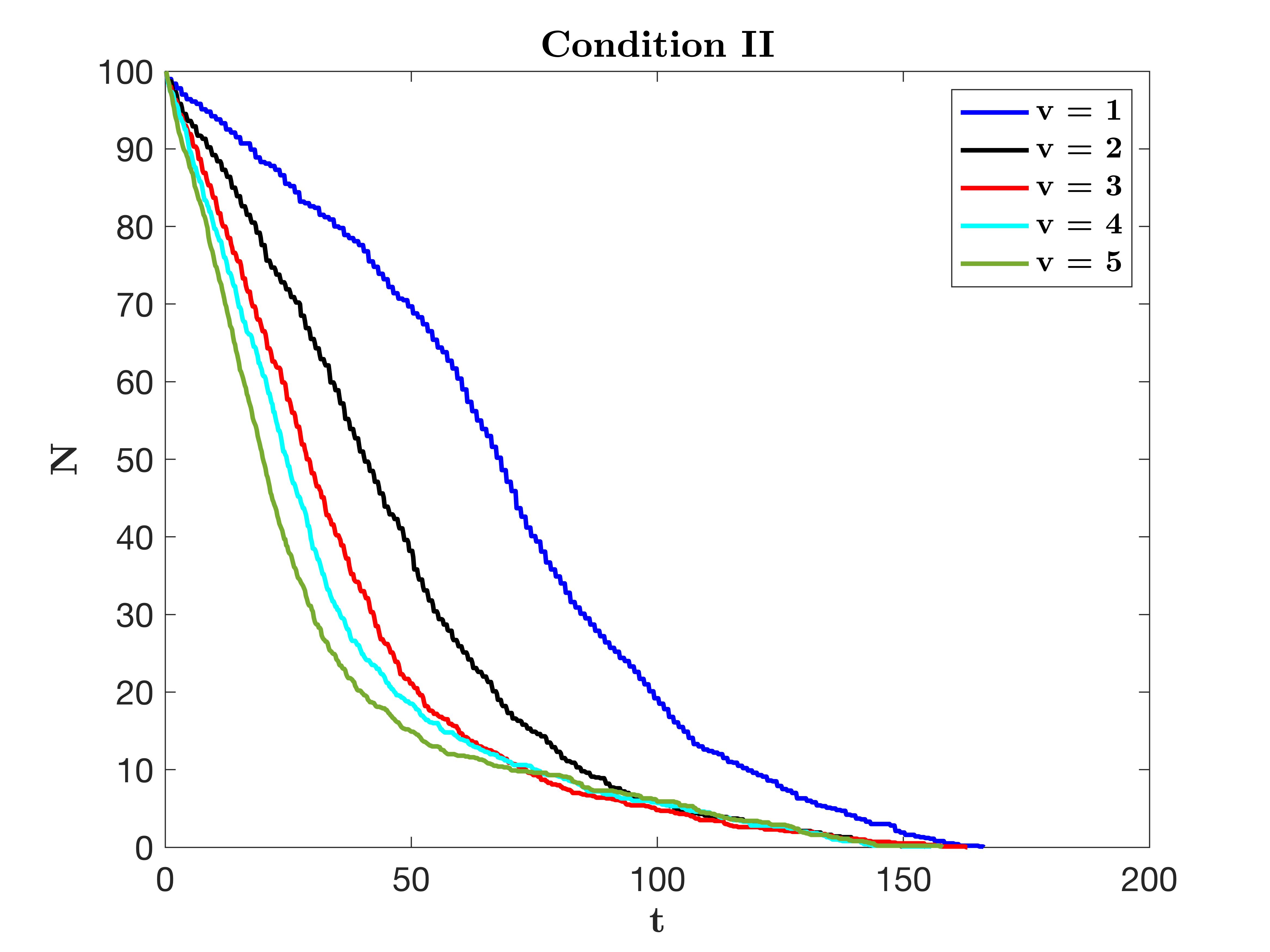}
\centering
\caption{\label{fig:epsart}Time evolution of the number of agents remained in the room in condition II.}
\end{figure}

\section{\label{sec:level1}Simulation experiments\protect}
\subsection{\label{sec:level2}Evacuation pattern and traffic analysis}
We conduct one-group-type simulations for individuals, 2-agent groups, 3-agent groups, 4-agent groups and 5-agent groups separately with a fixed total number of agents ($N(t=0)=480$). Initially, all agents are randomly distributed in the evacuation area as shown in Fig.1.
To study how the group structure affects PED when the full binding effects are included, we start with qualitative and quantitative analyses of the overall evacuation pattern. Since the waiting process is activated and multi-speed-agents are employed, parameters used in the simulation experiments, as shown in Table II,  are different from those shown in Table I.
\begin{table}[h!]
\centering
 \begin{tabular}{||c p{1cm}<{\centering} ||} 
 \hline
 Parameters &  Value \\ [1ex] 
 \hline
 $k_{S}$ & 8 \\ 
 $k_{S}'$ & 6 \\
 $k_{D_{0}}$ & 2 \\
 $k_{p}$ & 6 \\
 $k_{d}$ & 6 \\
 $\alpha $ & 0.5 \\
 $\delta$ & 0.1 \\
 $d_{l}$ & 4 \\
 $N_{l}$ & 2 \\
 $\theta_{err}$ & 0.2 \\ 
 $d_{0}$ & 3 \\ [1ex]
 \hline
 \end{tabular}
 \caption{Parameters settings for the simulation experiments.}
 \label{table:1}
\end{table}
Typical snapshots of the distribution of evacuating agents are shown in Fig. 7, from which three phases can be identified from the evacuation processes: (1) a movable phase shortly after the beginning when most agents can move in a relatively smooth manner; (2) a jammed phase following the movable phase when the most agents are getting obstructed; (3) and a rarefied phase in the end after the crowding has been released. As one can see, the emergence of three phases is universal among three simulation experiments with different group types, though the details of each are different.

\begin{figure}[htbp]

\begin{subfigure}{0.15\textwidth}
    \includegraphics[width=0.9\linewidth]{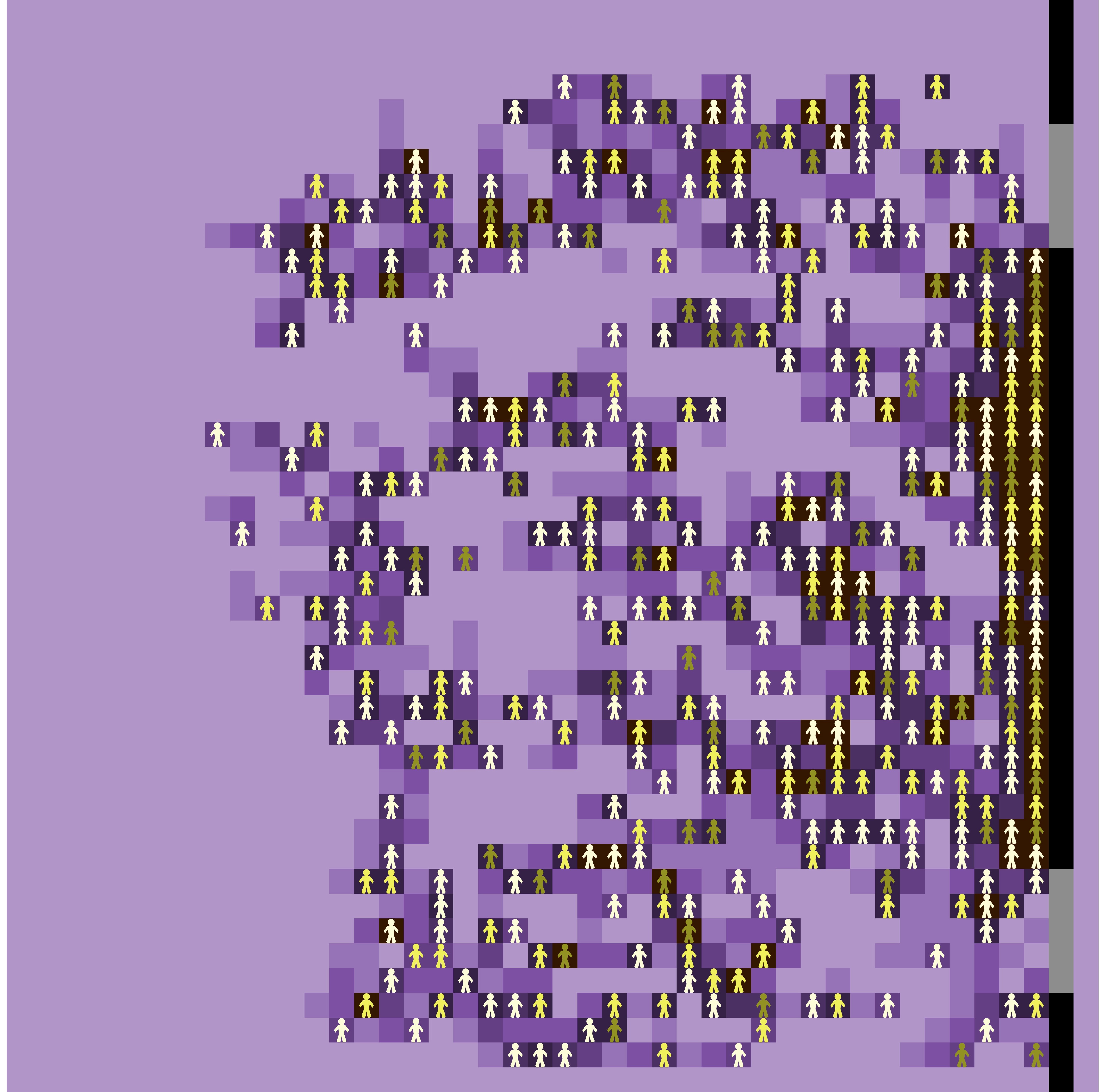}
    \subcaption{}
\end{subfigure}%
\begin{subfigure}{0.15\textwidth}
    \includegraphics[width=0.9\linewidth]{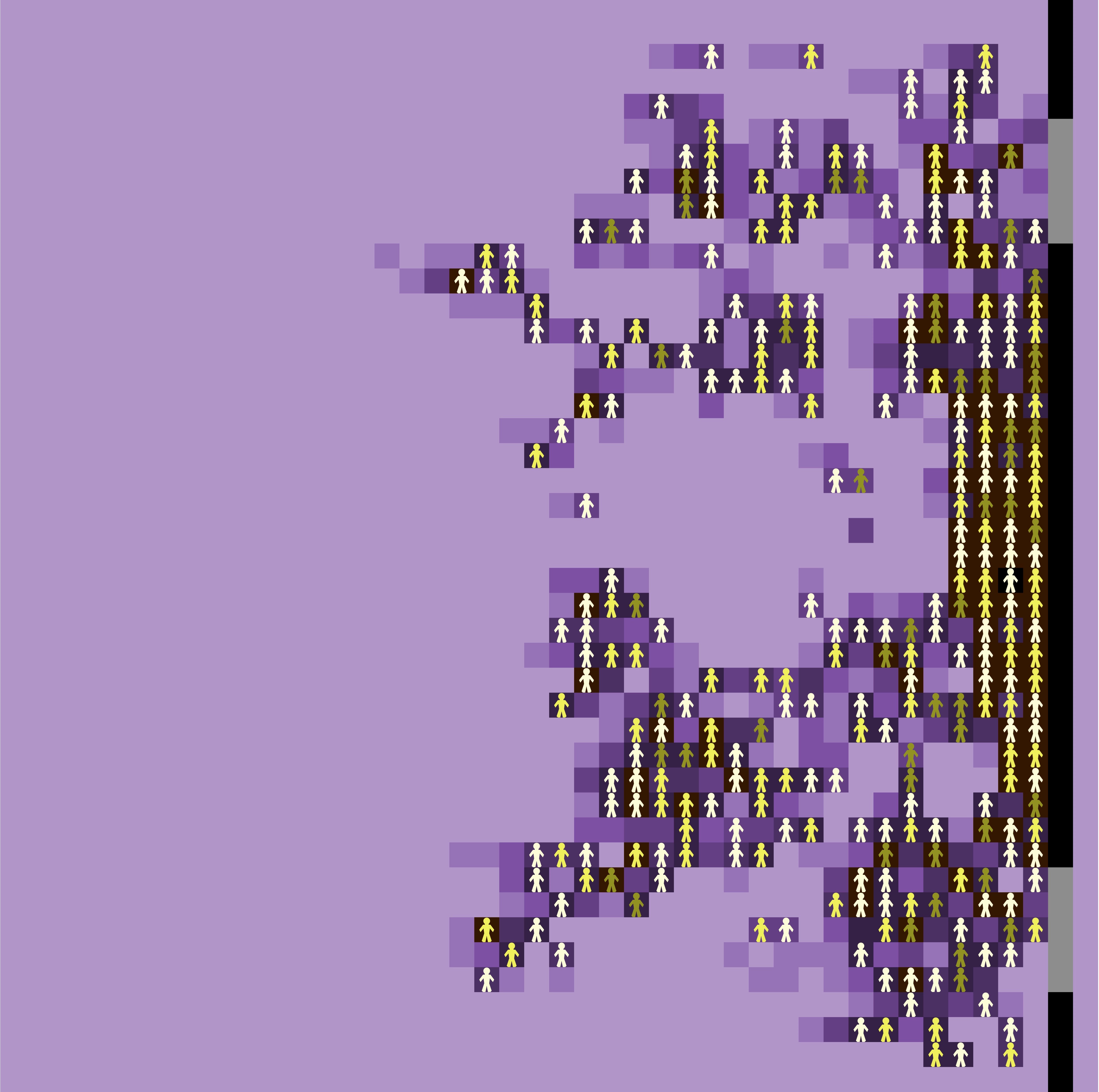}
    \caption{}
\end{subfigure}
\begin{subfigure}{0.15\textwidth}
    \includegraphics[width=0.9\linewidth]{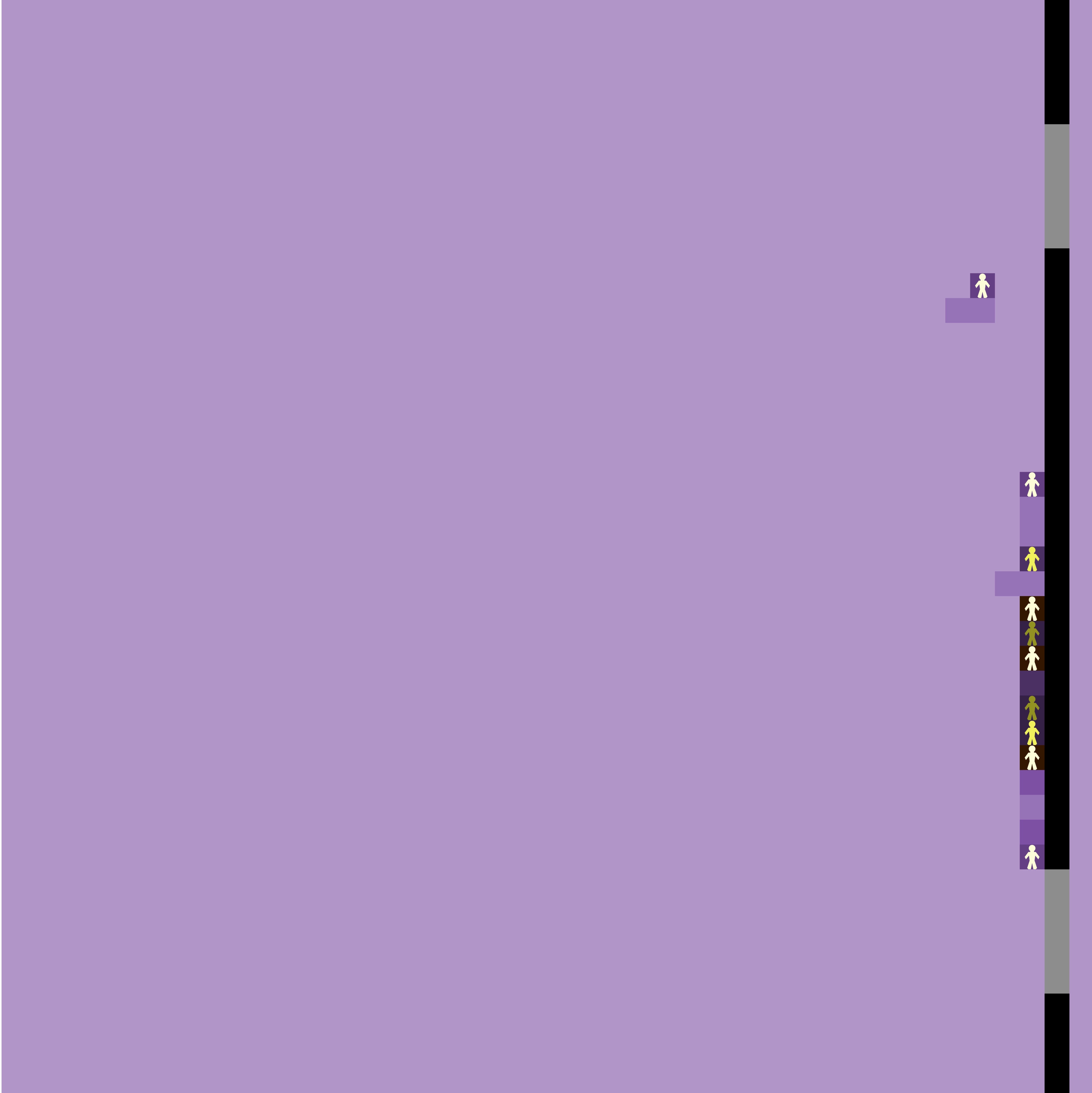}
    \caption{}
\end{subfigure}
\begin{subfigure}{0.15\textwidth}
    \includegraphics[width=0.9\linewidth]{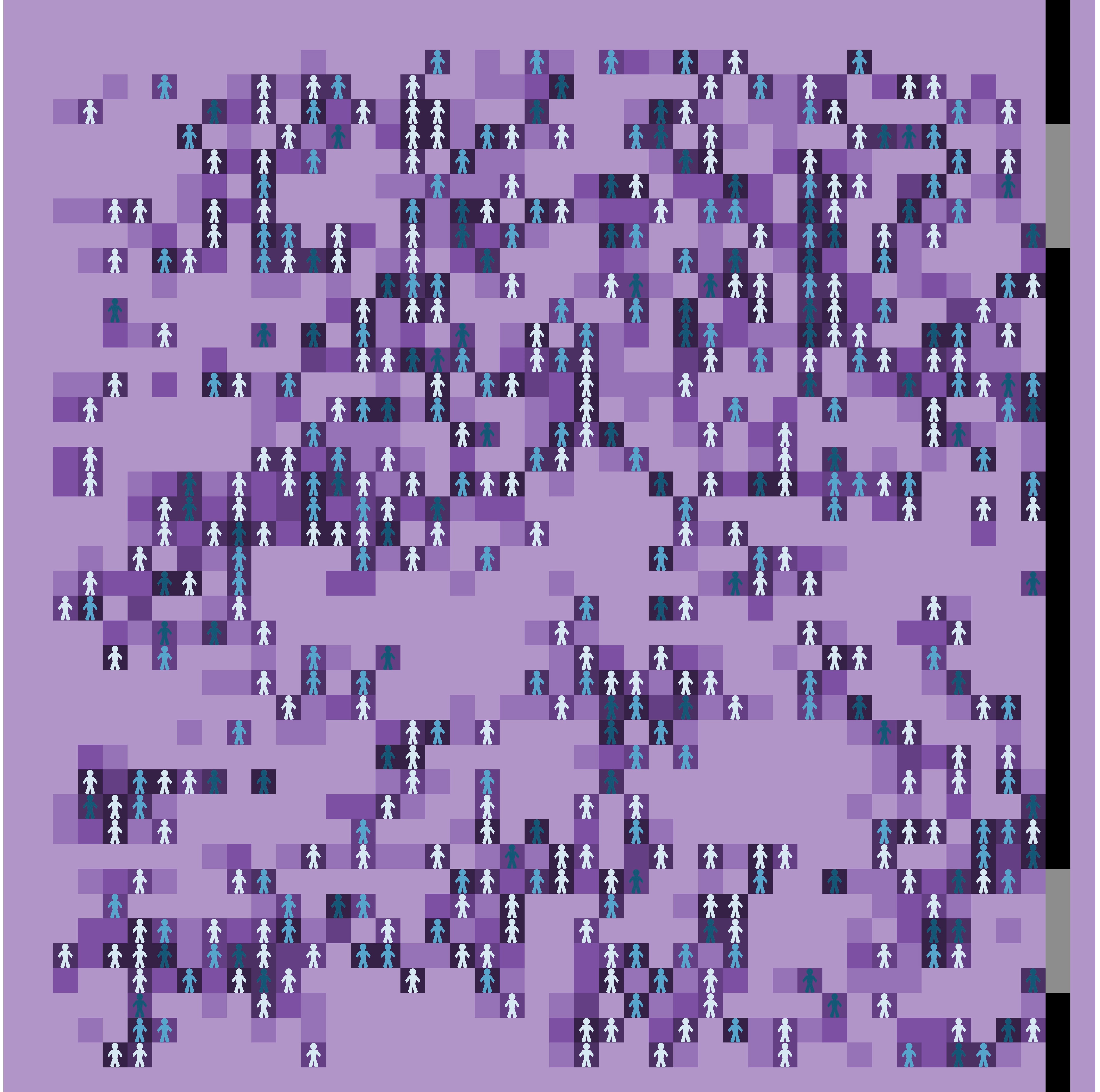}
    \subcaption{}
\end{subfigure}%
\begin{subfigure}{0.15\textwidth}
    \includegraphics[width=0.9\linewidth]{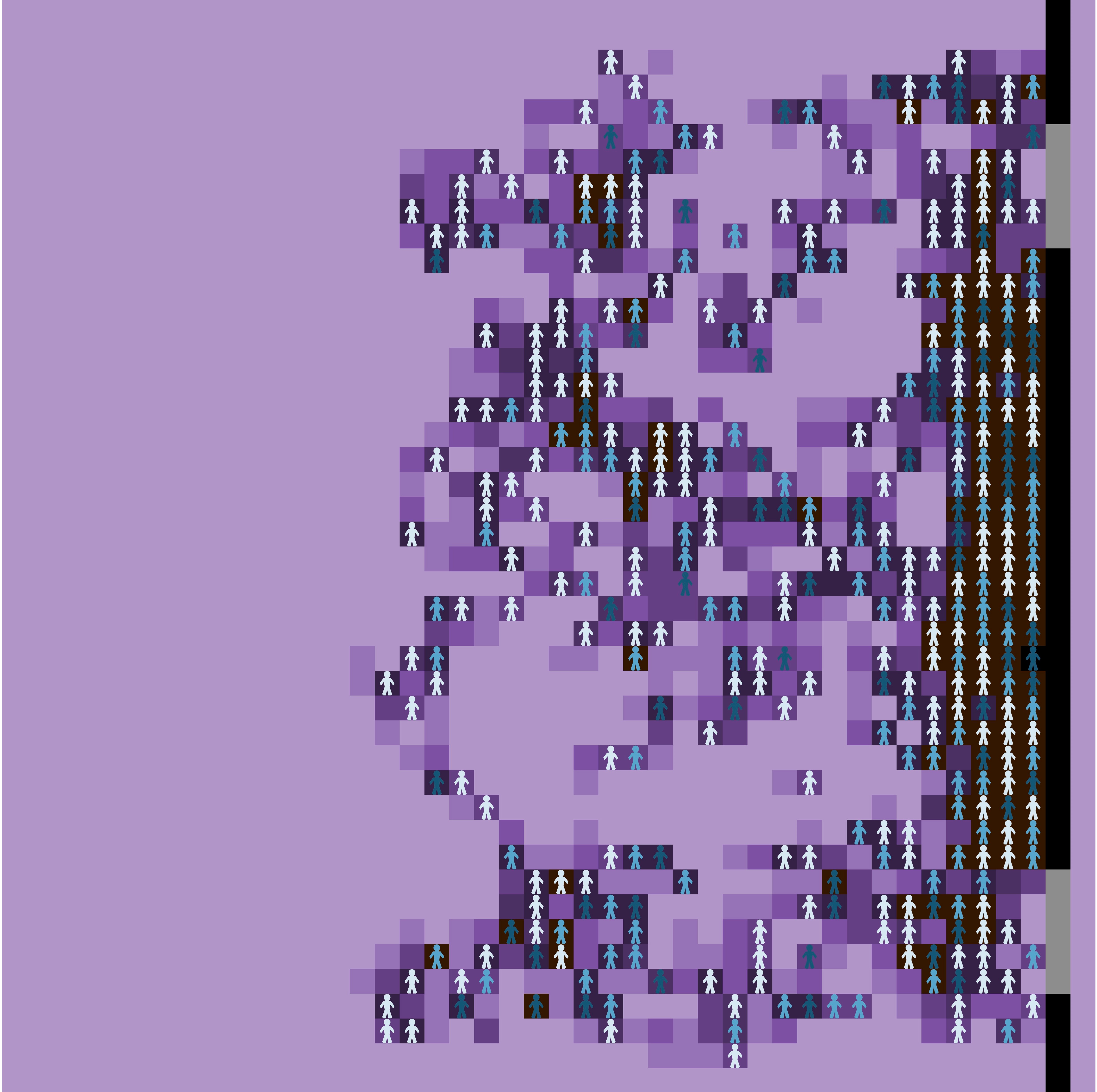}
    \caption{}
\end{subfigure}
\begin{subfigure}{0.15\textwidth}
    \includegraphics[width=0.9\linewidth]{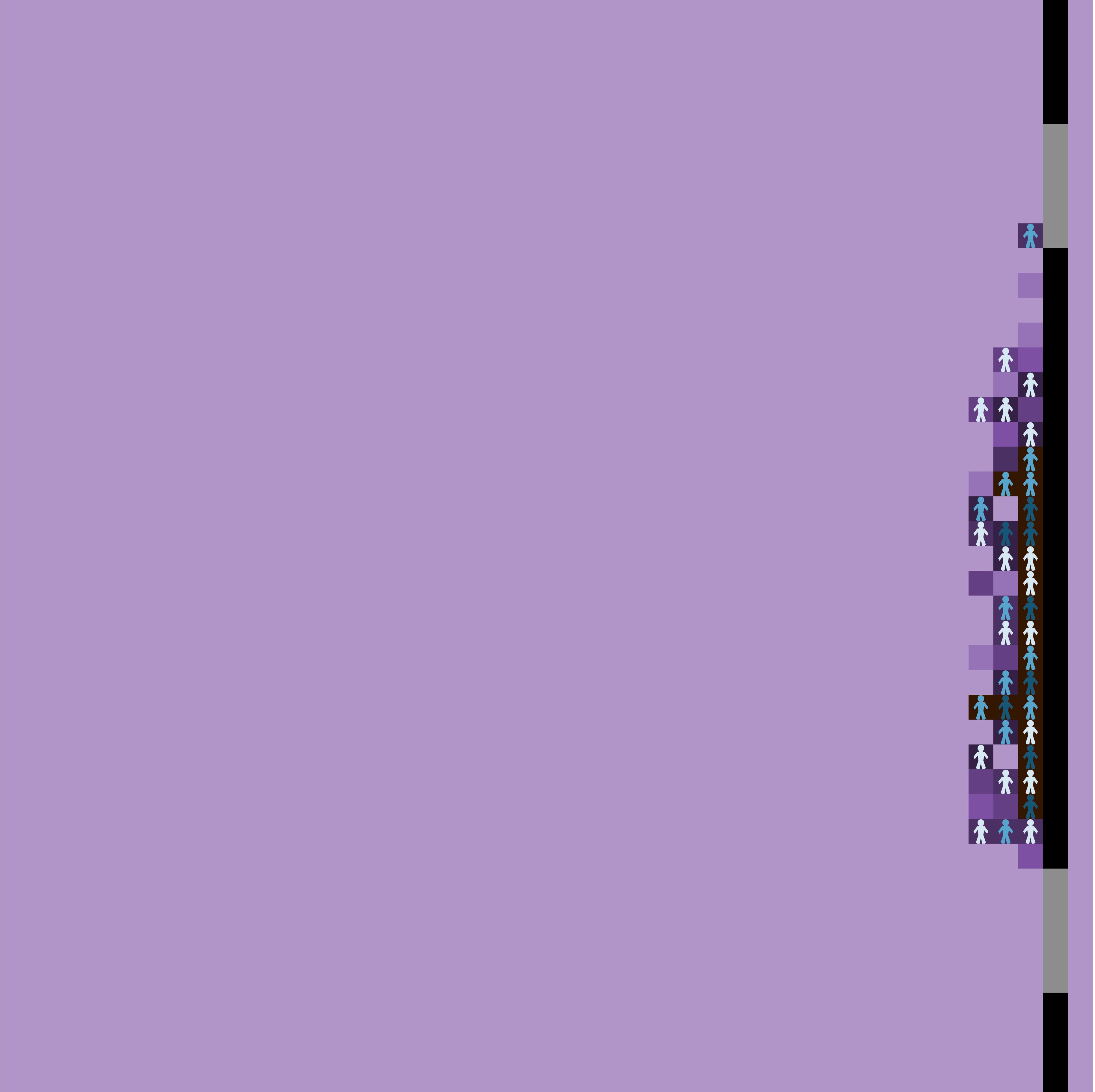}
    \caption{}
\end{subfigure}
\begin{subfigure}{0.15\textwidth}
    \includegraphics[width=0.9\linewidth]{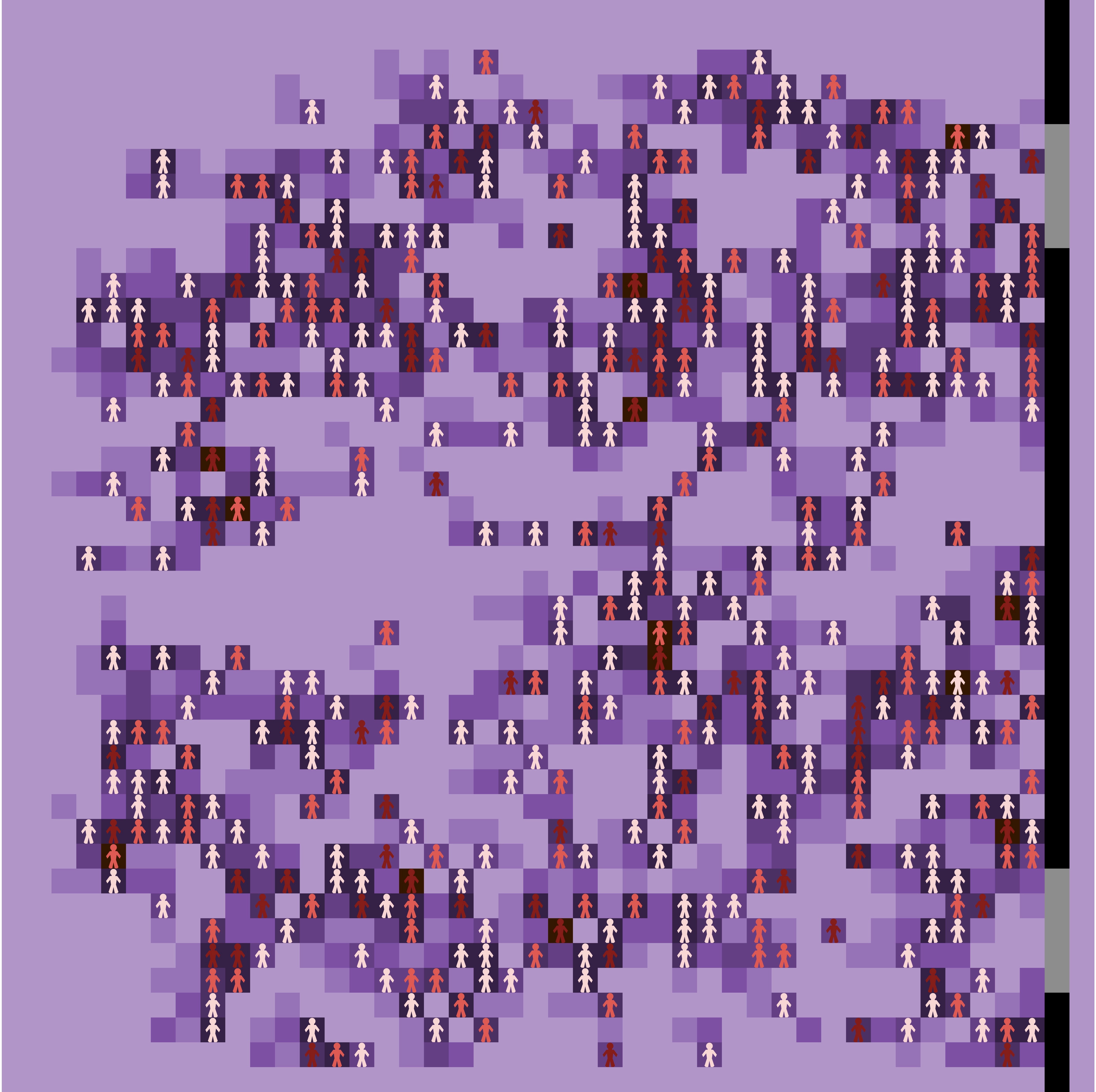}
    \subcaption{}
\end{subfigure}%
\begin{subfigure}{0.15\textwidth}
    \includegraphics[width=0.9\linewidth]{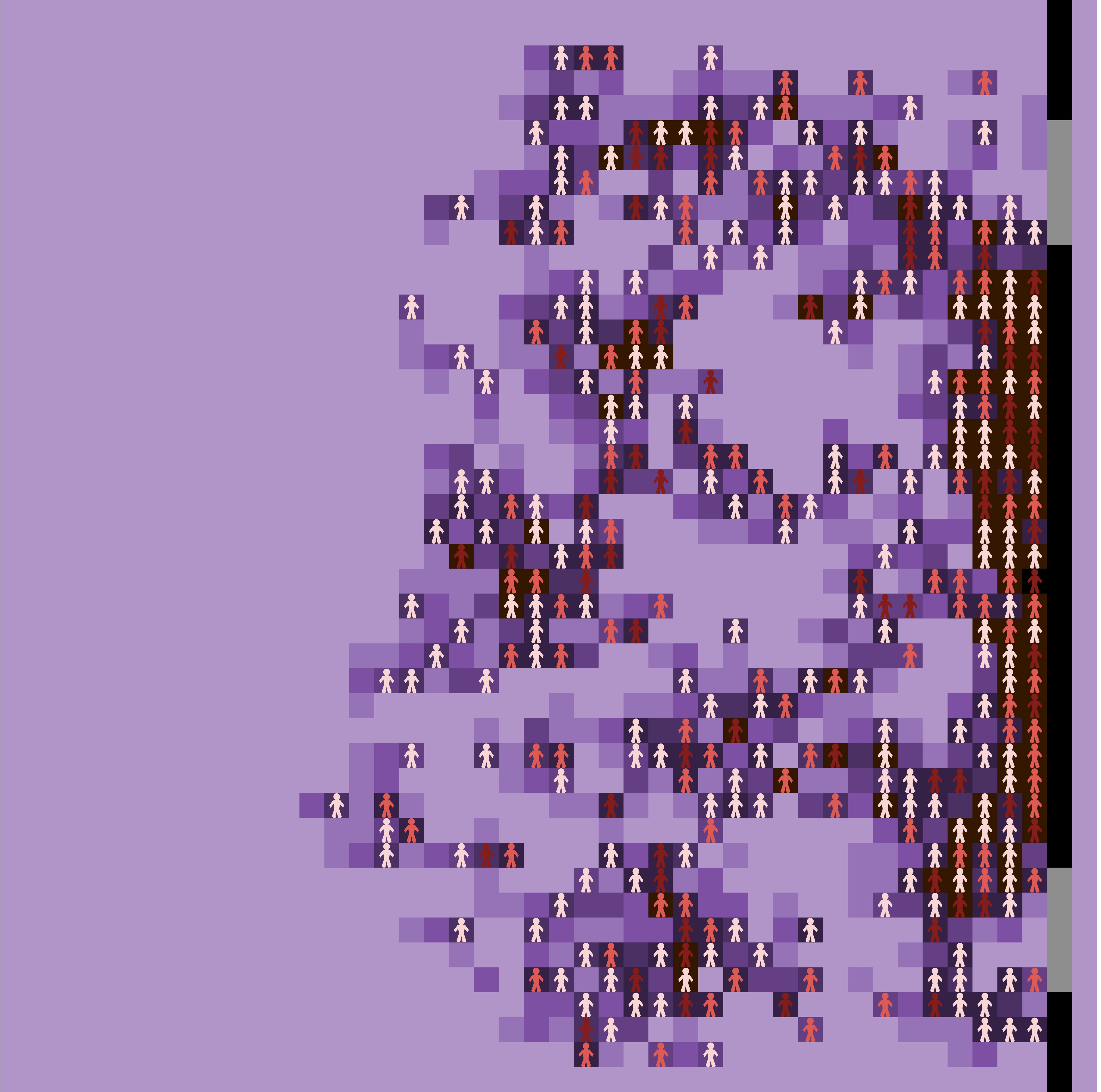}
    \caption{}
\end{subfigure}
\begin{subfigure}{0.15\textwidth}
    \includegraphics[width=0.9\linewidth]{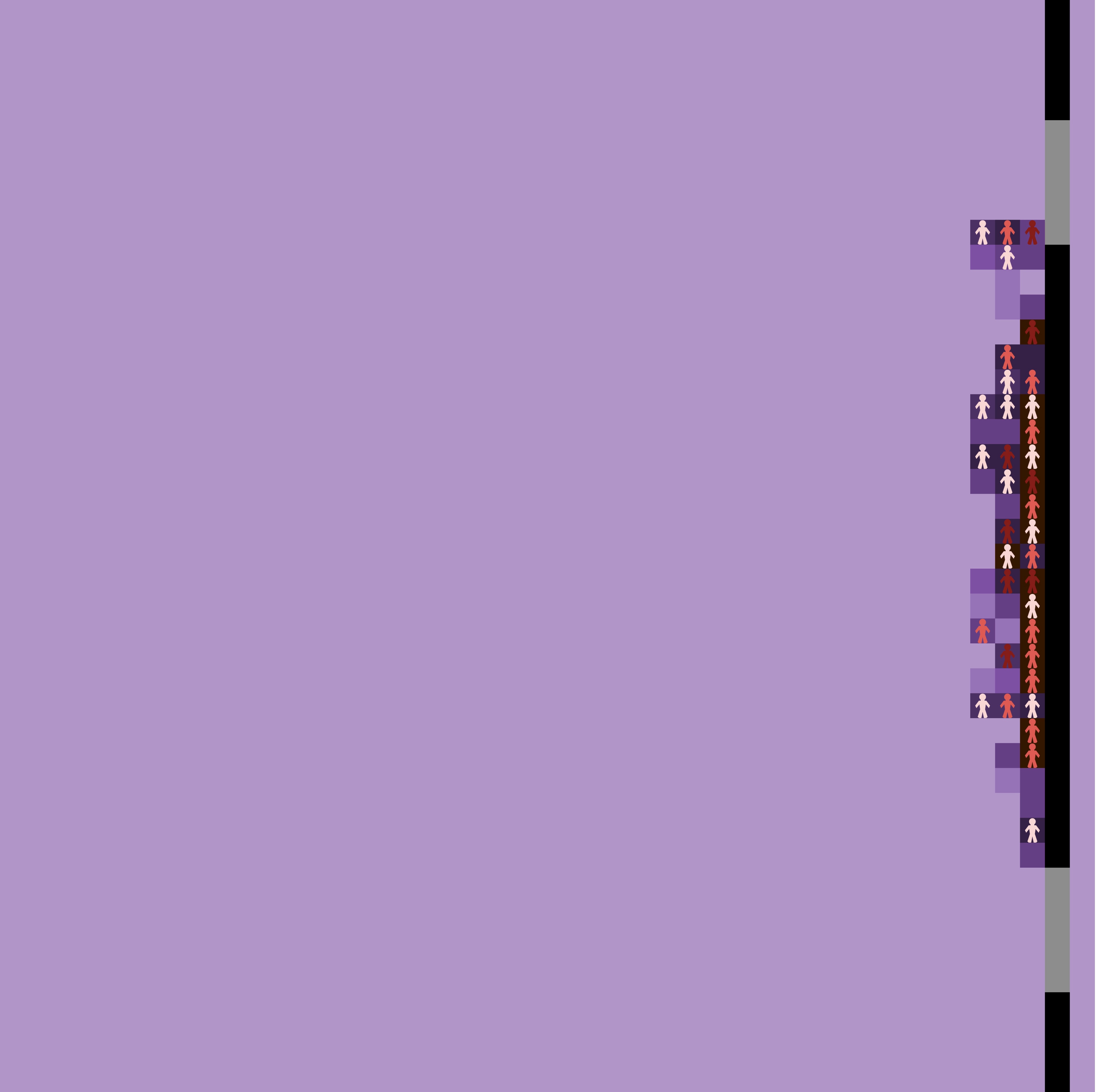}
    \caption{}
\end{subfigure}
\caption{\label{fig:epsart}Evacuation patterns for 2-agent groups (Fig.7(a-c)), 3-agent groups(Fig.7(d-f)), 4-agent groups(Fig.7(g-i)). Different patterns for the movable, jammed and rarefied phases are exhibited from left to right. The gradient of purple color in the background indicates the strength of the dynamic floor field. In the jammed phase, the dynamic floor-field dominates.}
\end{figure}

To quantify the level of jamming, we define a quantity of “traffic” to be \textit{the ratio of the number of moving agents moving to the total number of agents in the evacuation area at a given time step}. Time evolutions of traffic are shown in Fig.8, with which one can clearly measure the duration of the three phases. The  monotonically decreasing green line in the earlier stage represents the traffic change in the movable phase; as the jammed phase is reached, the almost horizontal red line shows that the traffic has been trapped in a very low level; whereas the monotonically increasing green line represents the entering into the rarefied phase where the jamming is gradually lifted and the evacuation is becoming smooth again. Note that, contrary to our intuition, the duration of jamming decreases monotonically with the increasing group size.

\begin{figure}[htbp]

\begin{subfigure}{0.23\textwidth}
    \includegraphics[width=1\linewidth]{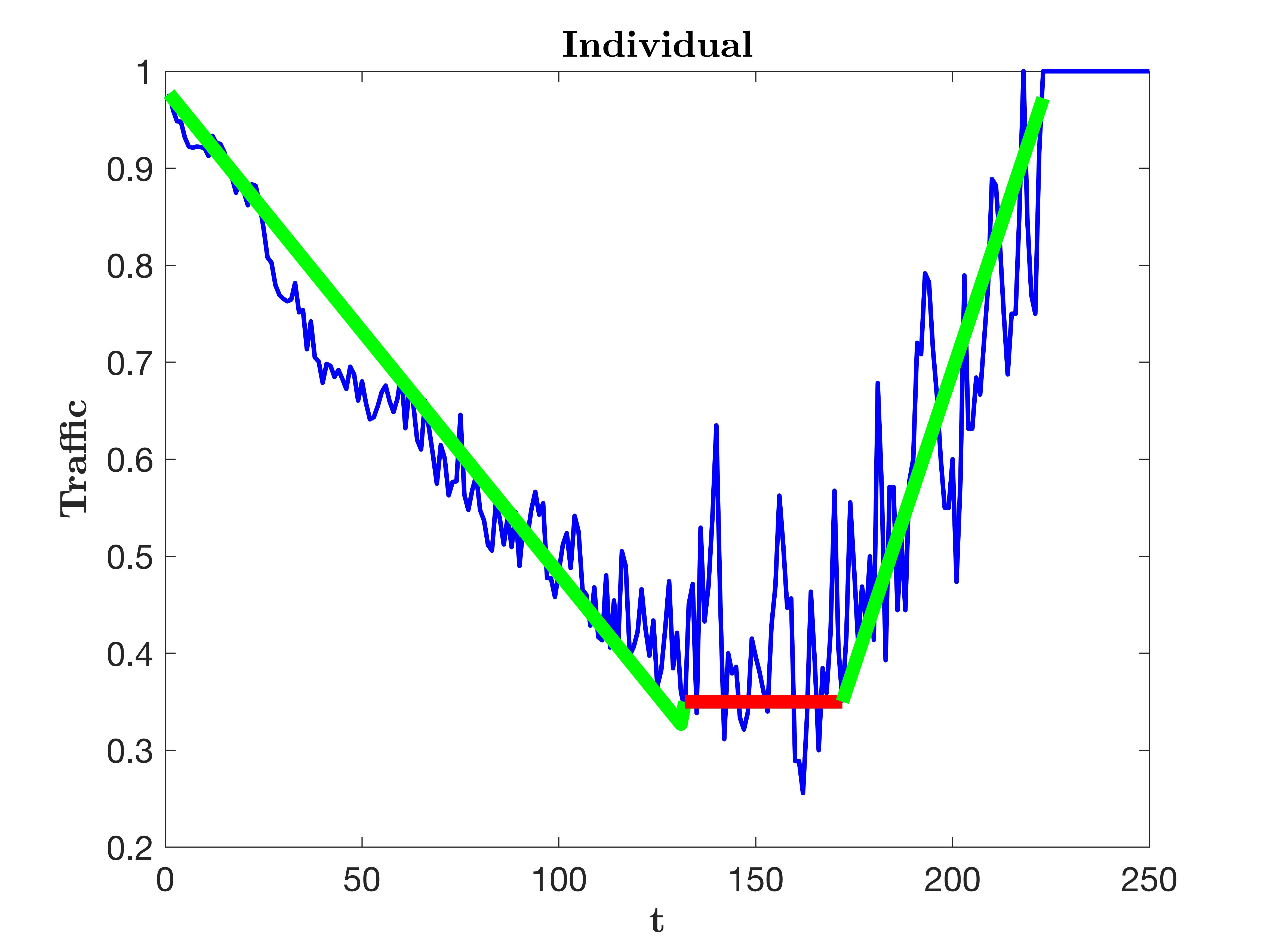}
    \subcaption{}
\end{subfigure}%
\begin{subfigure}{0.23\textwidth}
    \includegraphics[width=1\linewidth]{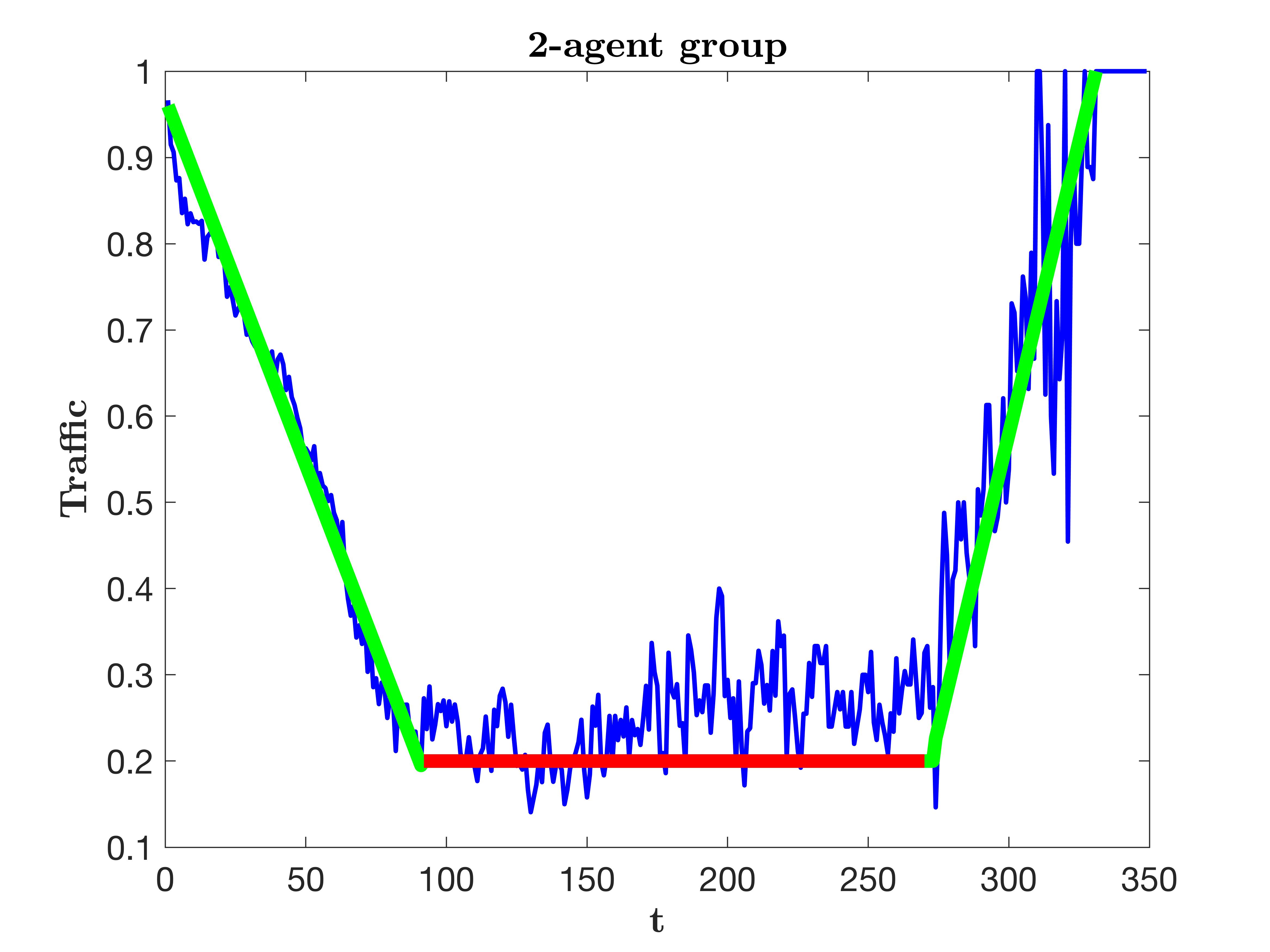}
    \subcaption{}
\end{subfigure} \quad
\begin{subfigure}{0.23\textwidth}
    \includegraphics[width=1\linewidth]{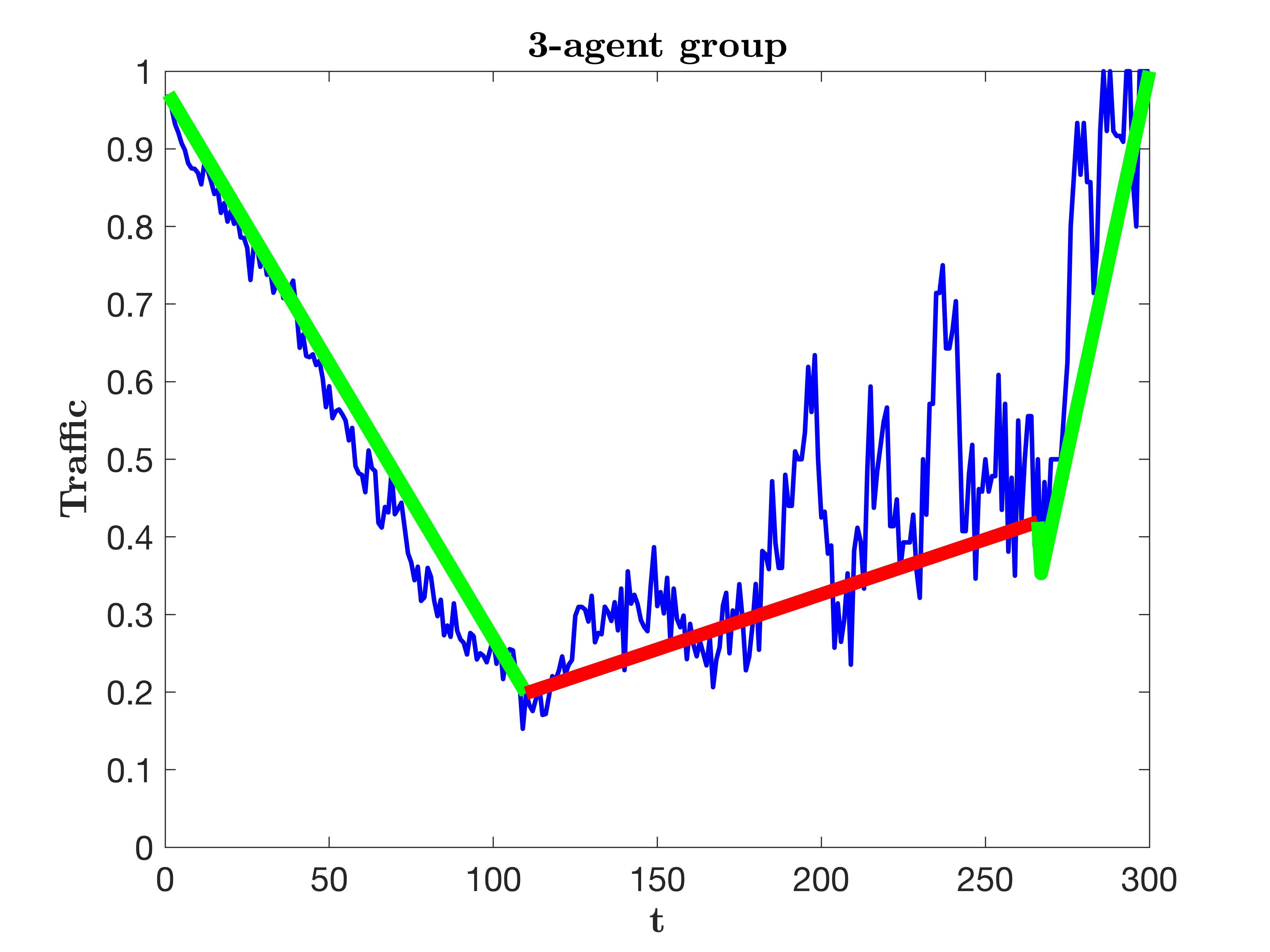}
    \subcaption{}
\end{subfigure}%
\begin{subfigure}{0.23\textwidth}
    \includegraphics[width=1\linewidth]{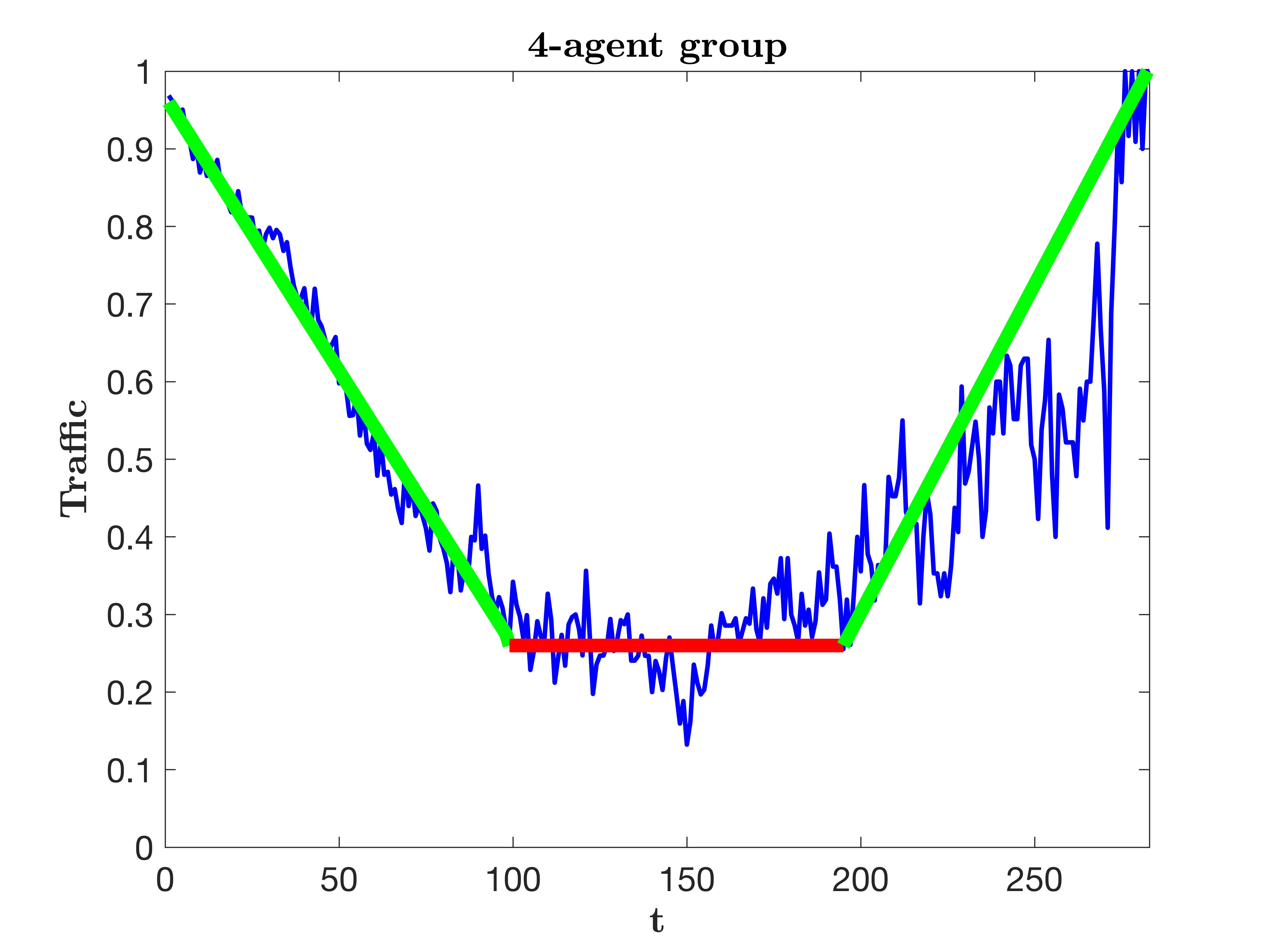}
    \subcaption{}
\end{subfigure} \quad
\begin{subfigure}{0.23\textwidth}
    \includegraphics[width=1\linewidth]{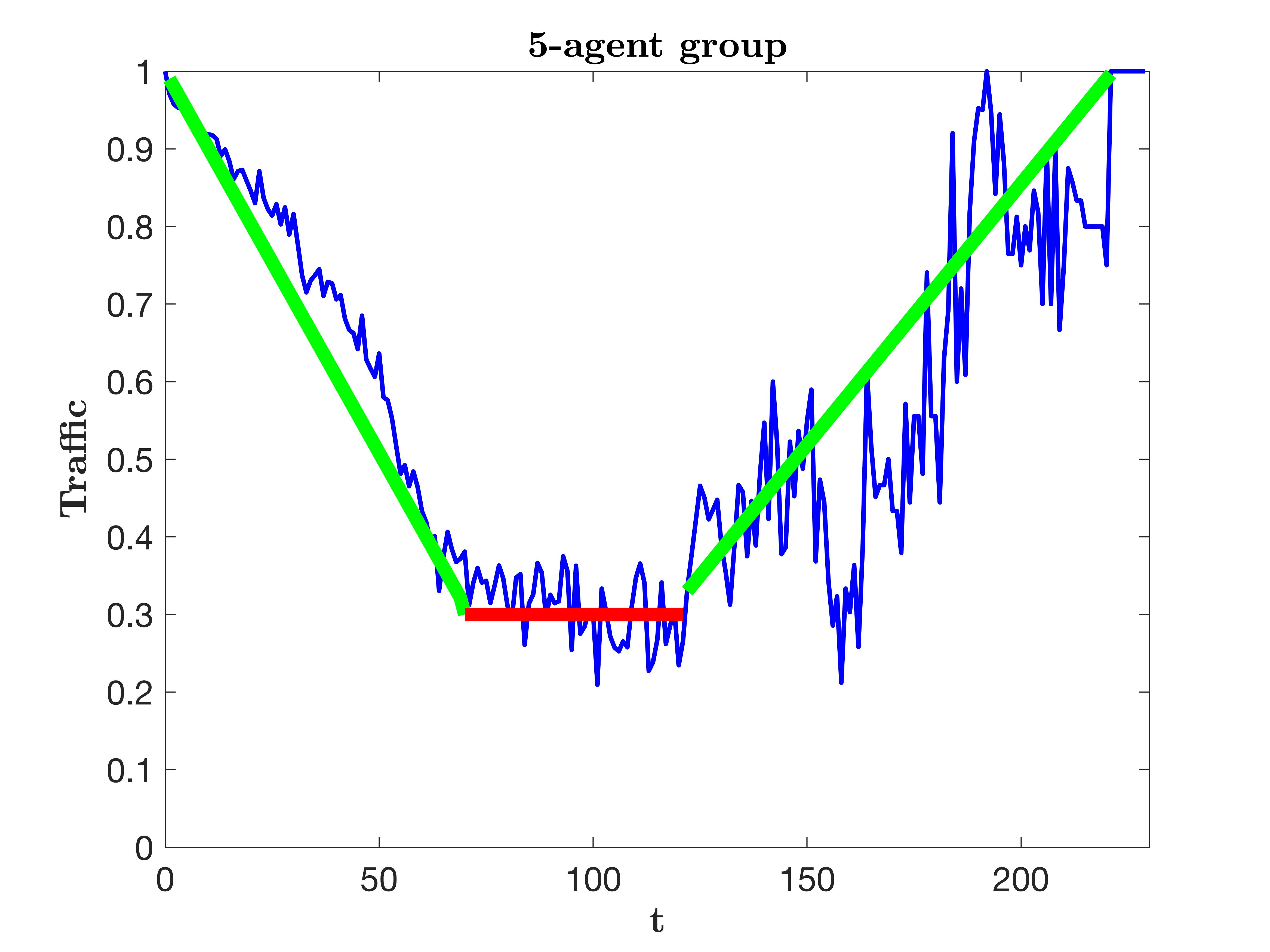}
    \subcaption{}
\end{subfigure}%
\caption{\label{fig:epsart}Time evolutions of traffic. For each type of group, durations of the  movable (time span of the left green line), jammed (time span of the middle red line) and rarefied phases (time span of the right green line) can be identified through drawing approximately the gradient of traffic changes.}
\end{figure}

\subsubsection{Negative dependence of evacuation time on group size}
Here we conduct three case studies, namely the complete binding case, the incomplete binding case and the no-binding case, about the total evacuation time ($T_{tot}$) and average evacuation time ($\bar{T}$) as defined in Sec.II.A. Note that complete binding means that, as is mentioned in Sec.II.E, followers will follow their group leaders and leaders will wait for their group followers; the incomplete binding refers to the lead-following mechanism proposed in the work of Lili Lu et al without the waiting mechanism for leaders\cite{lu_study_2017}. In addition, we use the case of no-binding as a reference for comparisons.

\begin{figure}[htbp]

\begin{subfigure}{0.4\textwidth}
    \centering
    \includegraphics[width=0.8\linewidth]{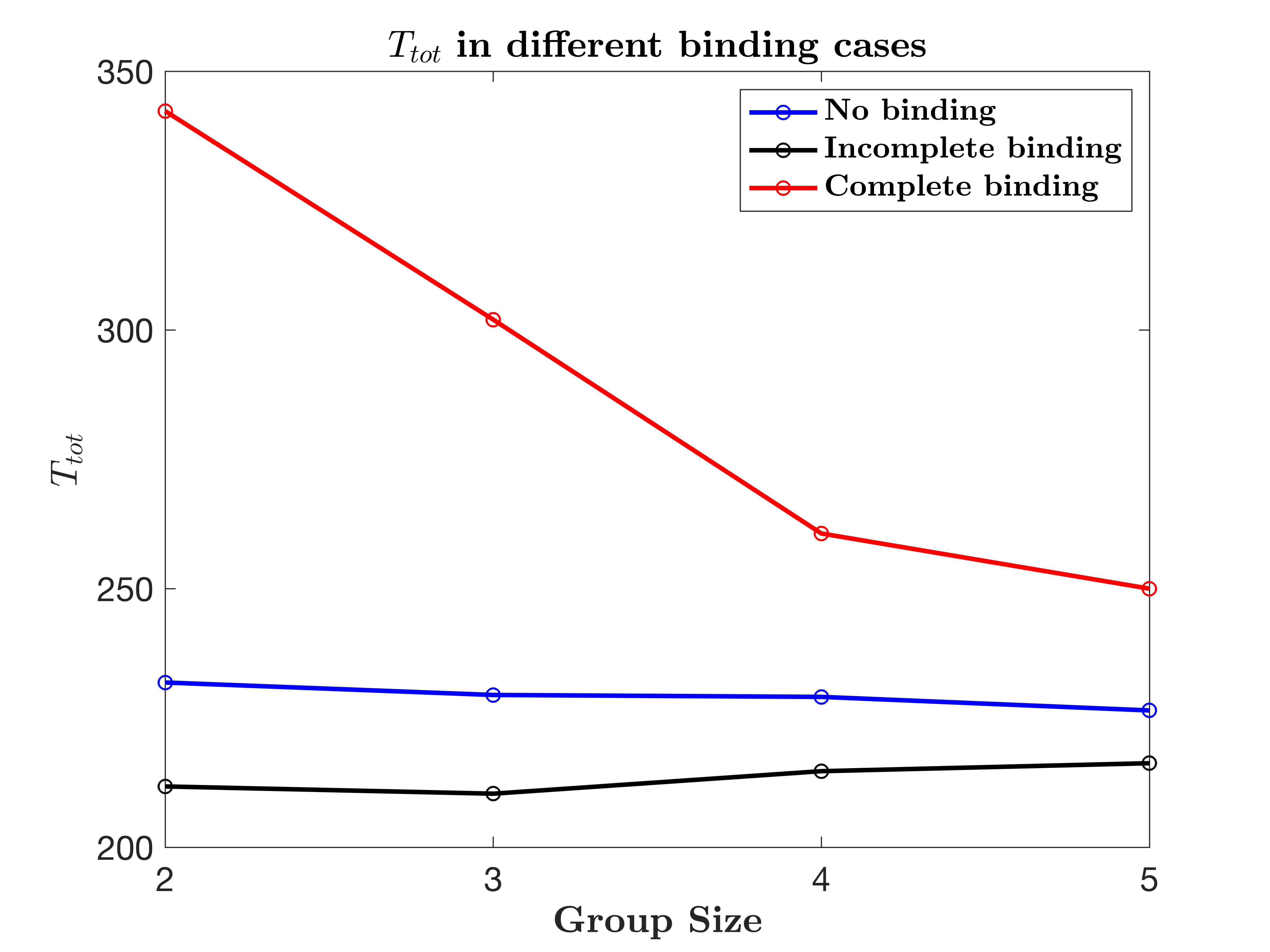}
    \subcaption{}
\end{subfigure} \quad
\begin{subfigure}{0.4\textwidth}
    \centering
    \includegraphics[width=0.8\linewidth]{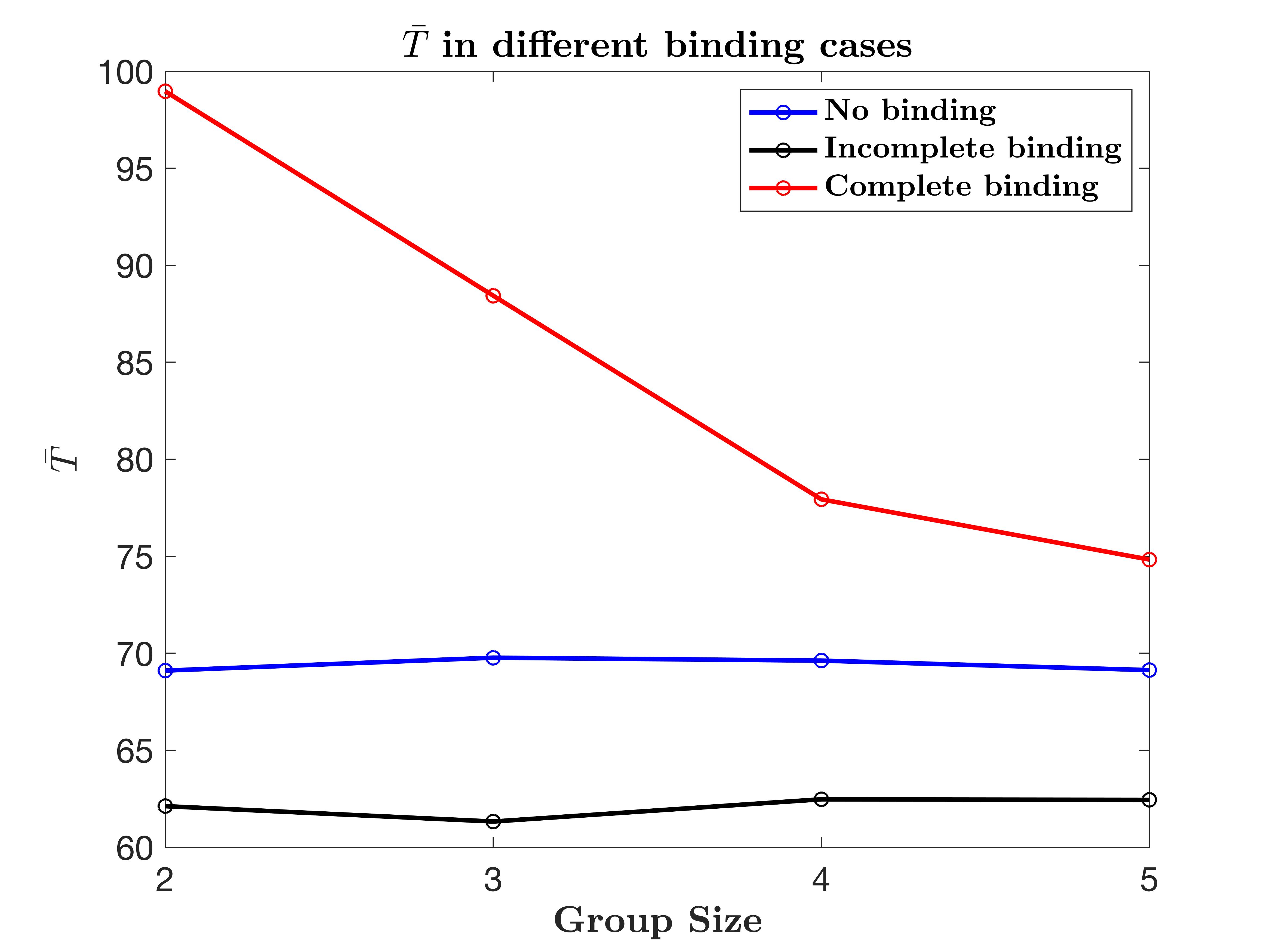}
    \subcaption{}
\end{subfigure}
\caption{\label{fig:epsart}Total and average evacuation times versus group sizes. For three different binding cases,  results of total evacuation time are shown in (a), and results of average evacuation time in (b).}
\end{figure}

As the results shown in Fig.9, while both the total and the average evacuation time become longer in the complete binding case, the correlation of these two quantities on the group size is negative. For the incomplete binding case, such a negative dependence turns into a weakly positive one. If the binding effects are completely removed (as shown in the no binding case), the correlation between the evacuation time and the group size is gone.

\begin{figure}[htbp!]

\begin{subfigure}{0.4\textwidth}
    \centering
    \includegraphics[width=0.8\linewidth]{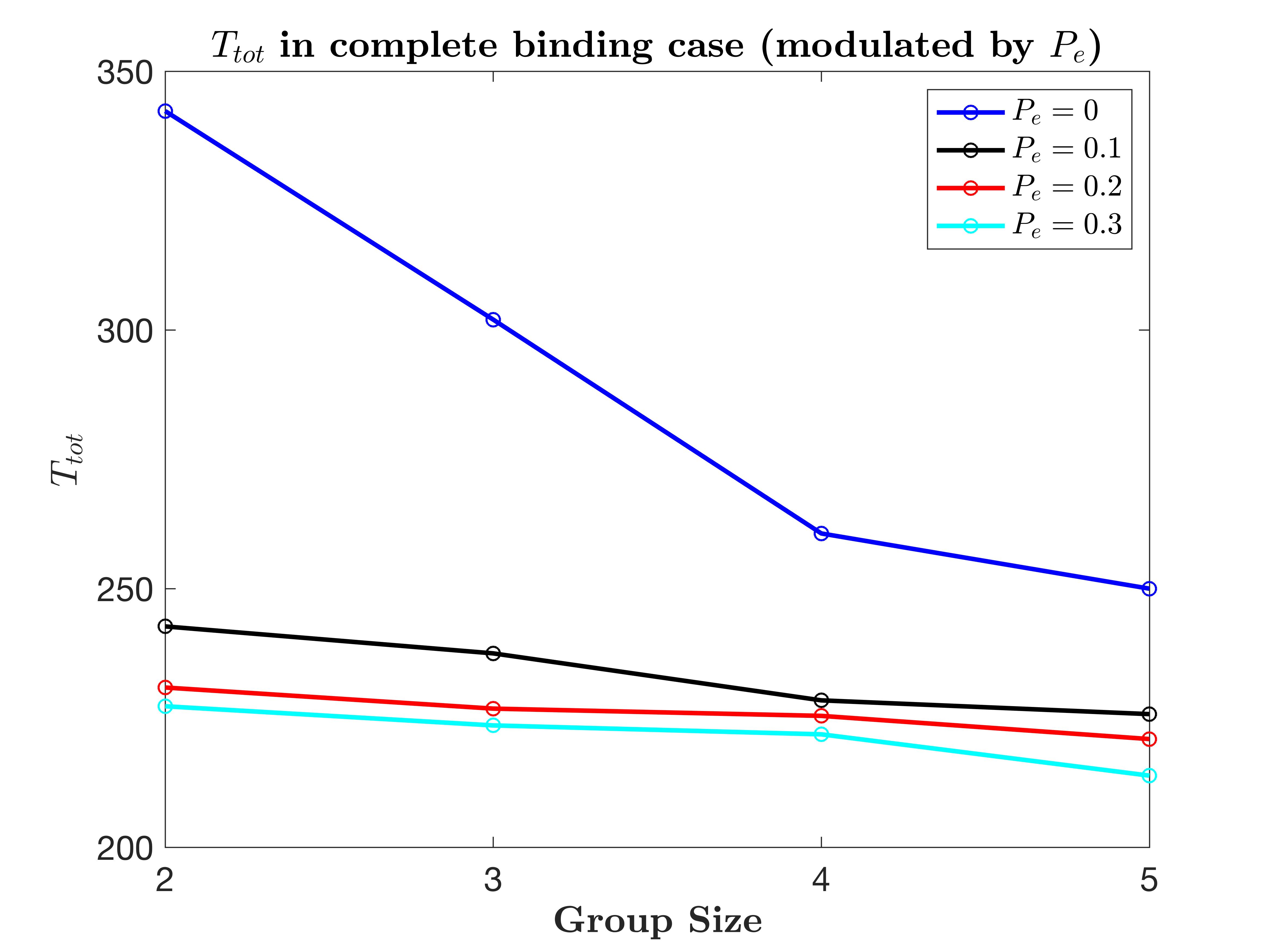}
    \subcaption{}
\end{subfigure}\quad
\begin{subfigure}{0.4\textwidth}
    \centering
    \includegraphics[width=0.8\linewidth]{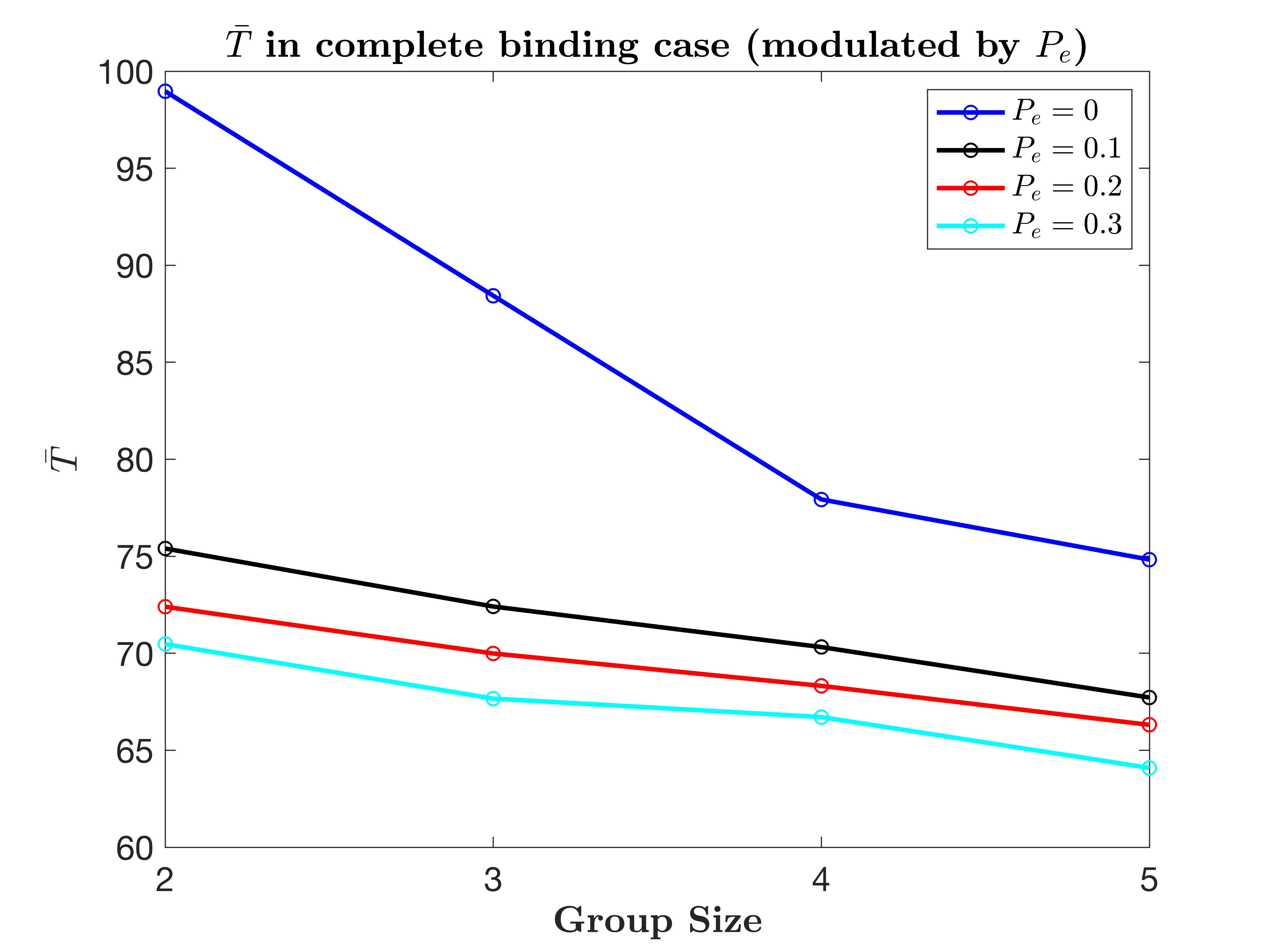}
    \subcaption{}
\end{subfigure}\quad
\begin{subfigure}{0.4\textwidth}
    \centering
    \includegraphics[width=0.8\linewidth]{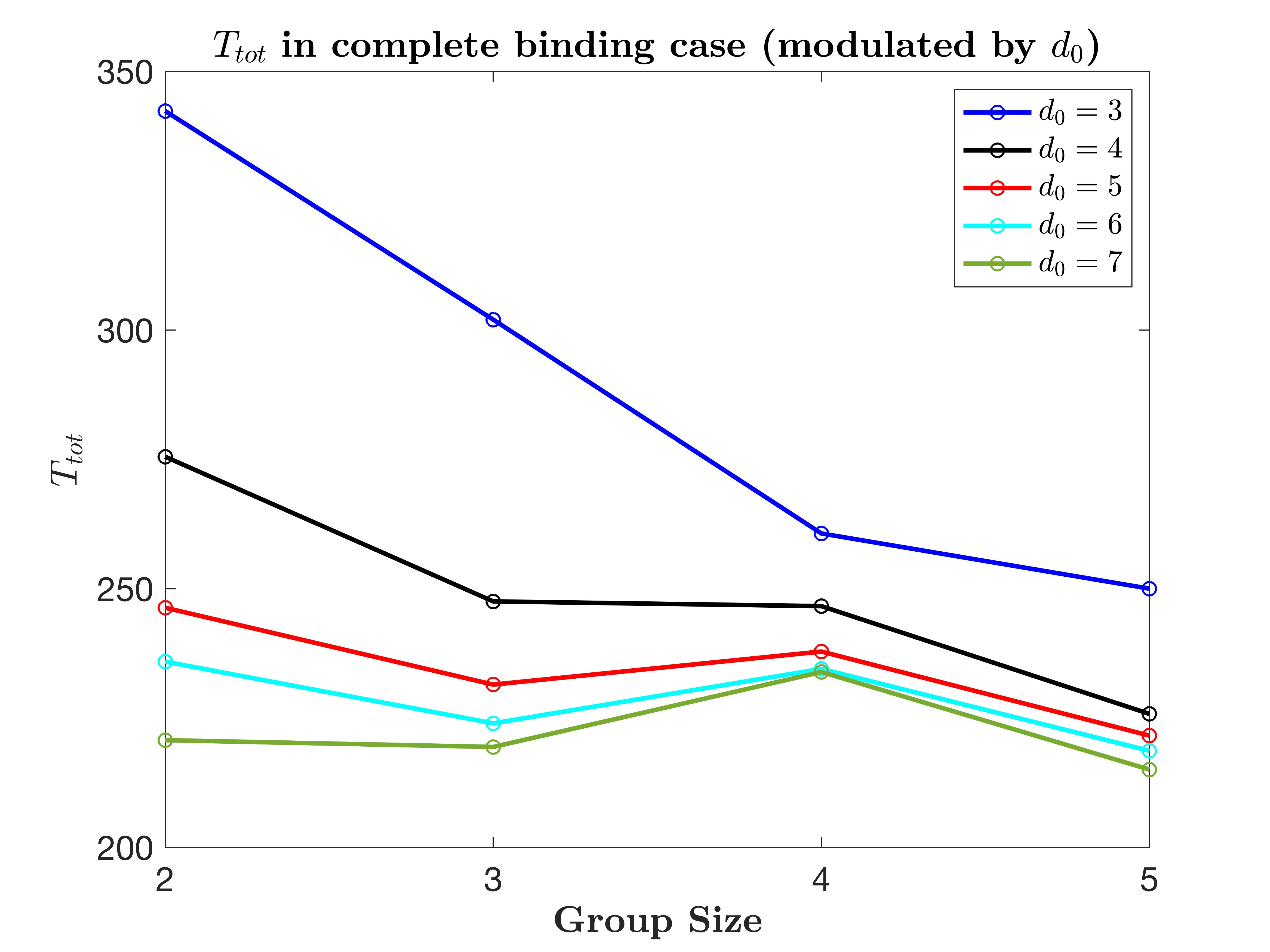}
    \subcaption{}
\end{subfigure}\quad
\begin{subfigure}{0.4\textwidth}
    \centering
    \includegraphics[width=0.8\linewidth]{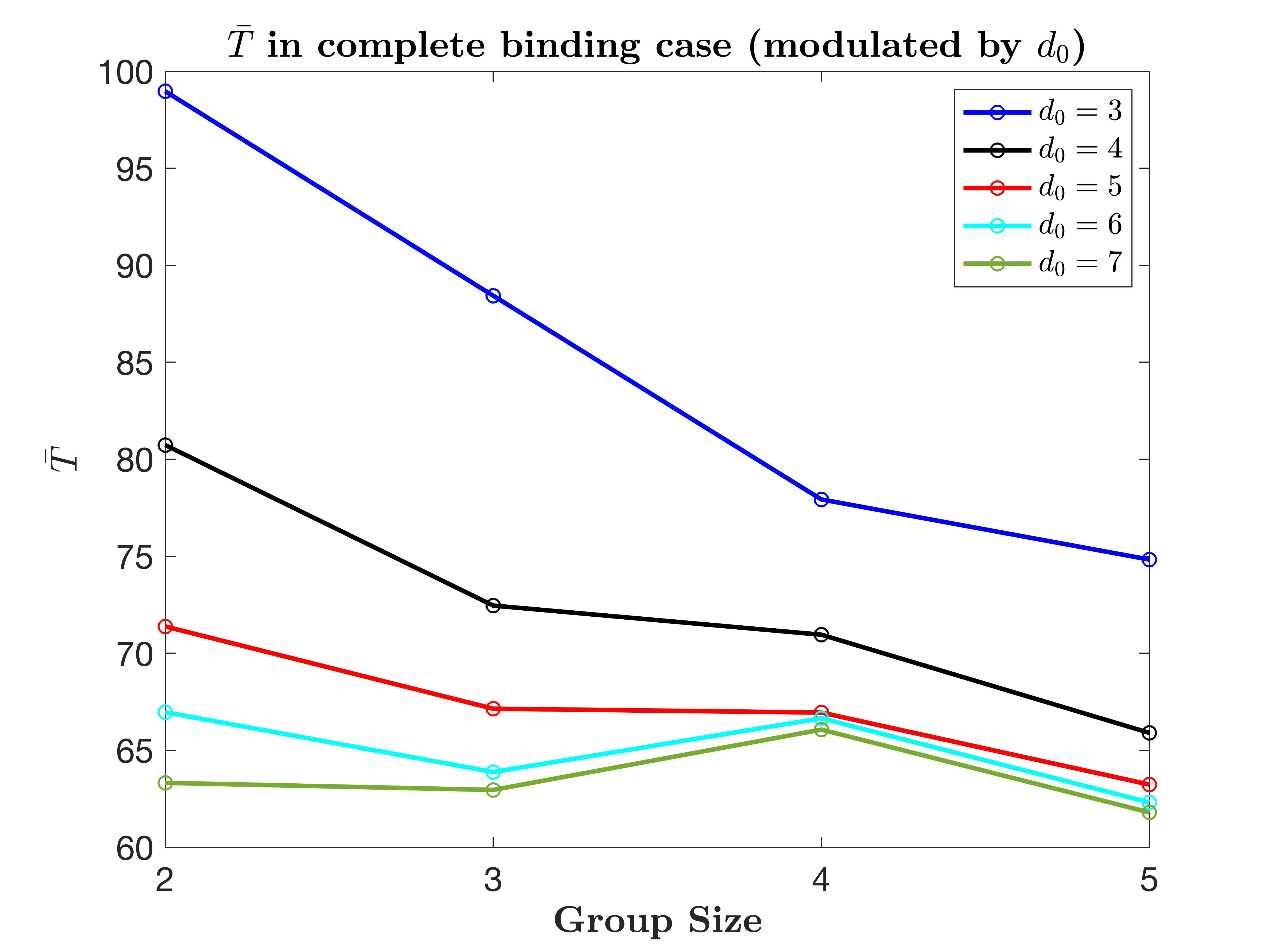}
    \subcaption{}
\end{subfigure}\quad
\caption{\label{fig:epsart}Total and average evacuation time versus group size under different strengths (modulated by $P_{e}$ in (a) and by $d_{0}$ in (b) of complete binding.}
\end{figure}
In reference to the no-binding case, the waiting mechanism in the complete binding makes it necessary for leaders, who have the highest mobility, to stop frequently and wait for other group members, a behavior that certainly leads to an increase in the overall evacuation time. In contrast, as there is only the following mechanism in the incomplete binding case, group members who have lower mobility can be attracted by the more speedy leaders who act like more rational guides. Hence chances of blocking together with other slower members driven by the dynamic floor-field are reduced. When the leaders are not to be dragged by slower members through the waiting mechanism, the evacuation process in the incomplete binding case becomes the fastest one.
It is worth further investigating the negative dependence of evacuation time on group size found in the case of complete binding. To this end, we adjust the strength of complete binding in the following ways:
(1) Set the probability for the failure of waiting mechanism to be $P_e$, such that even if the critical condition of waiting is satisfied, there is still a probability of $P_e$ that the leader will “make a mistake” and choose not to wait.
(2)  Relax the fixed distance $d_0$ for the waiting mechanism to a flexible one.
Note even we change values of $P_e$ and $d_0$, the mechanisms of  waiting and following still work together so that the binding is still complete, though the “strength” of binding is modulated.
The results of adjusting the binding strength by changing $P_e$ are shown in Figs. 8(a) and 8(b), and the results of adjusting the binding strength by changing $d_0$ are shown in Figs. 8(c) and 8(d). The negative dependence of evacuation time on group size diminishes as the binding strength is reduced. The overall decrease in evacuation time as either $P_e$ increases or $d_0$ increases can be attributed to the fact that fewer leaders’ are waiting for their group members. 

\section{Mixing index}
By far we have made a qualitative argument that the negative dependence of evacuation time on group size emerges only when the complete binding effects are included. And we have observed that the more the strength of complete binding is reduced the weaker the negative dependence is. However, we need to have a quantitative understanding about how the  complete binding contributes to such a negative dependence.

\subsection{Definition of the mixing index}
As shown in Sec.III.A, we find that the duration of the jammed phase differs for different group sizes. Thus we may propose an ansatz that the mixing of members from different groups could be the main reason for such differences. In the model design, there is a critical distance $d_0$ over which the leader must stop and wait until the rest of the group members get closer under the influence of the following mechanism. However, if the system resides in a state where different group members are highly mixed, grids next to a group member would be occupied by members from other groups with a high probability, making it difficult for this agent to get close to his/her leader. On a larger scale, the mixing of different group members can make the distances between leaders and their farthest group members more likely to exceed $d_0$, hence triggering the waiting more frequently. 
The reasoning on the elongation of duration for the jammed phase in a highly mixed state under the complete binding mechanism inspires us to construct an entropy-like quantity to describe the degree of mixing. Before defining, we need to make a representation transformation from agents to grids. When constructing the model, it is a natural choice for us to focus on the agents. However, it would be more convenient to take an Eulerian view when constructing a spatially local state variable. We define a unit cell, which, as shown Fig. 11, consists of 9 Moore neighboring grids. In the center grid i of this unit cell, by analogy with the definition of entropy in statistical mechanics, we define a local quantity $M_i$, called the local mixing index hereafter. 
\begin{figure}[htbp]
\includegraphics[width=8cm]{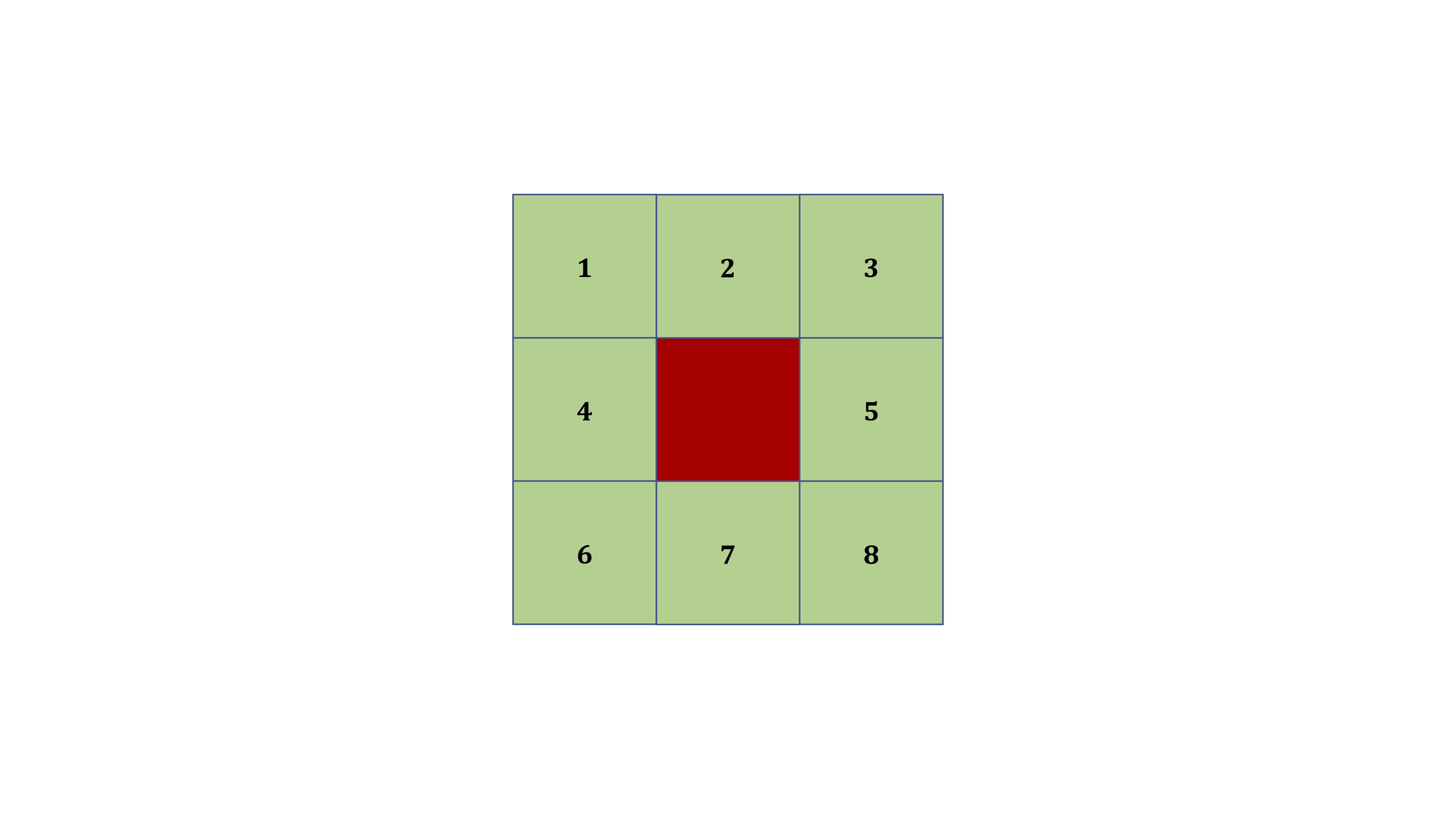}
\centering
\caption{\label{fig:epsart}A unit cell for the calculation of mixing index. The center grid is colored in red and the periphery grids are colored in green. For the evacuation area configured in Sec.II.A, there are 1600 such unit cells in total.}
\end{figure}
For a unit cell with centring grid i and neighboring grids $j=1,...,8$, the mixing index is defined as,
\begin{equation}
M_i=\gamma_i n_i ln(\sum_j n_j+1).
\end{equation}
where binary variables $n_i$ and $n_j$ indicate the existence or absence of agents in the grids, and i is a coefficient for the alignment of group indices for the existing agents,
\begin{equation}
\gamma_i = \prod_j' \delta(g_i-g_j)
\end{equation}
where $g_i$ and $g_j$  are the group indices for agents staying in the center and in the neighbors. The prime on the factorial operator $\prod_j '$ indicates that the components of products are limited to those grids that are occupied. Hence as long as the center grid and at least one of the periphery grid are occupied by the agents who belongs to the same group, $i=0$ or else $i=1$. The constant 1 in the logarithmic operator is for the avoidance of divergence of the mixing index, if all the neighboring grids are empty. Note that our mixing index will reach its maximum if an agent is trapped by agents from different groups, while it will reach its minimum if the agent is surrounded by its own group members, or else simply if the grid is empty. Lastly, for those unit cells that have centers on the boundary, we may simply extend the periphery grids out of the boundary, and set those grids as unoccupied ones.

The mixing index for the whole evacuation system can be obtained by summing the local mixing indices,
\begin{equation}
M=\sum_i M_i
\end{equation}
As the system mixing index $M$ will increase monotonically as the local mixing level of different groups increases, together with the distribution of local mixing indices, it can pave the road for the upcoming quantitative analysis of the negative dependence of evacuation time on the group size.

\subsection{Analysis with the mixing index}
Firstly, time evolution of the system mixing index is shown in Fig.12(a) for a 2-agent group evacuation simulation with complete binding (parameters are set to be the same as those shown in Sec.III.B). We may observe that the system stays in a highly mixed state in the initial stage. As the degree of mixing decreases almost monotonically along with the evolving evacuation process, the total evacuation time should have a dependence on the maximum value of the mixing index. Hereafter, we use $M_{max}$ to be the maximal system mixing index during one run of the evacuation simulation.

\begin{figure}[htbp]

\begin{subfigure}{0.23\textwidth}
    \includegraphics[width=0.9\linewidth]{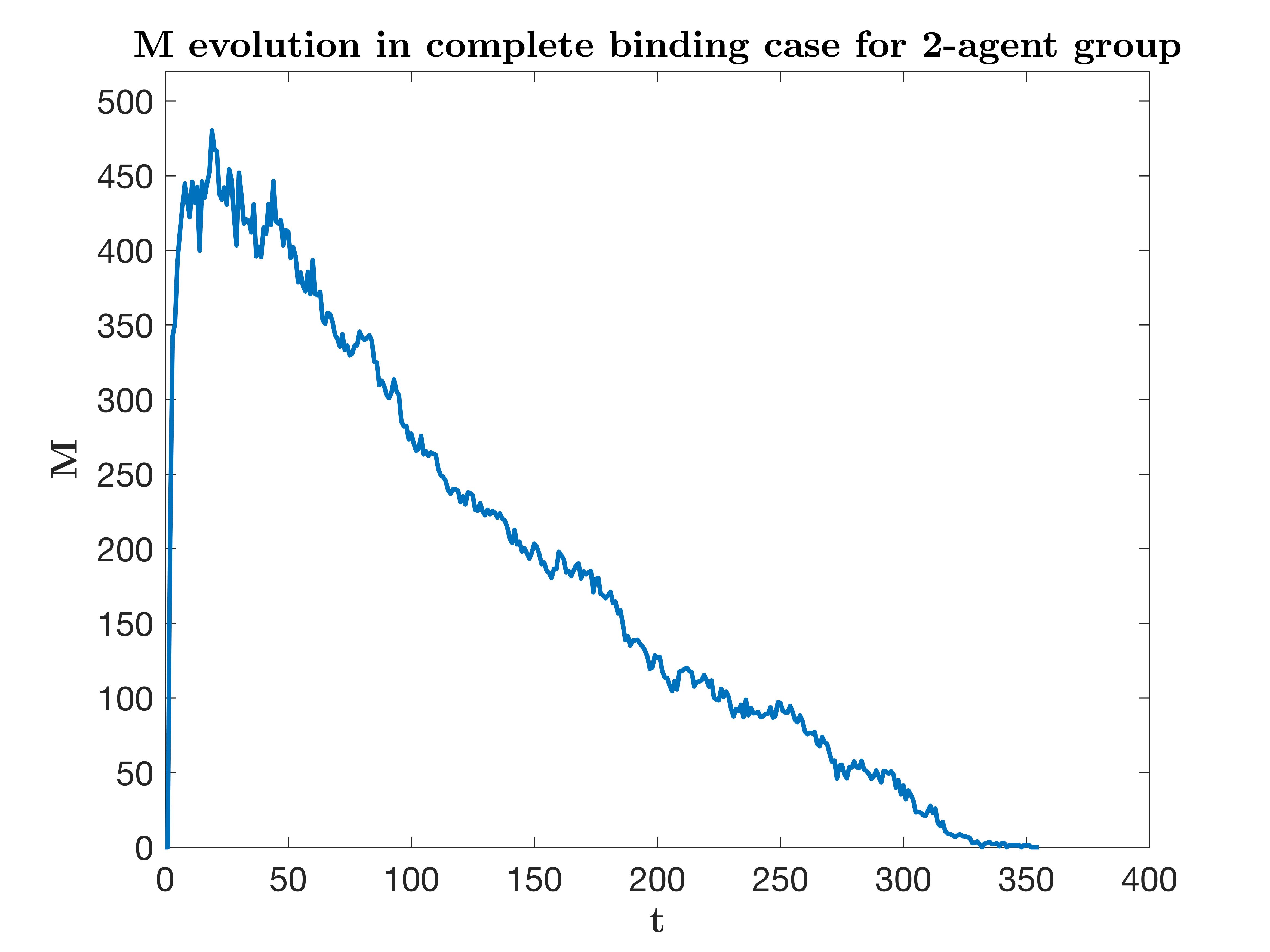}
    \subcaption{}
\end{subfigure} \quad
\begin{subfigure}{0.23\textwidth}
    \includegraphics[width=0.9\linewidth]{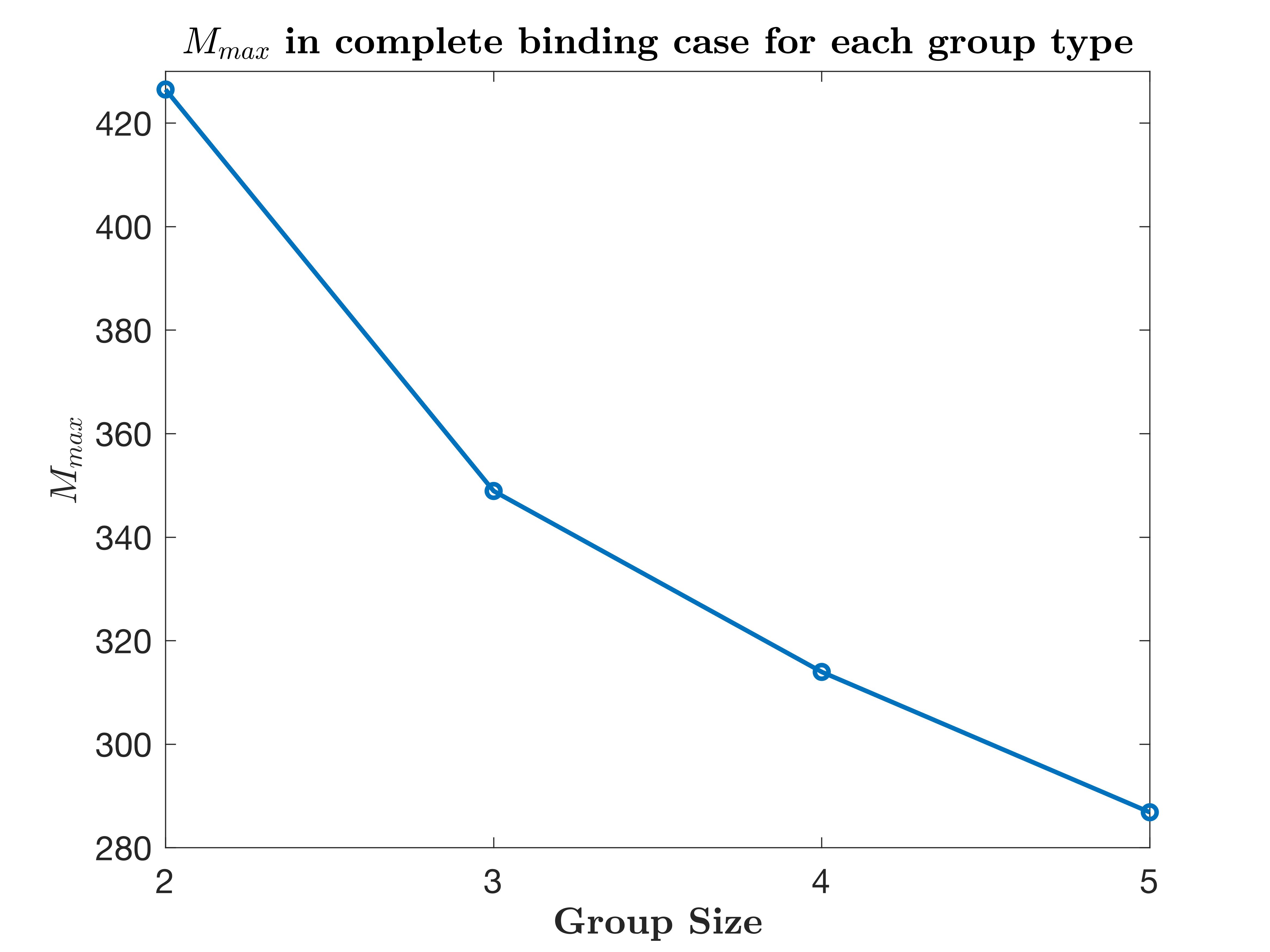}
    \subcaption{}
\end{subfigure} 
\begin{subfigure}{0.23\textwidth}
    \includegraphics[width=0.9\linewidth]{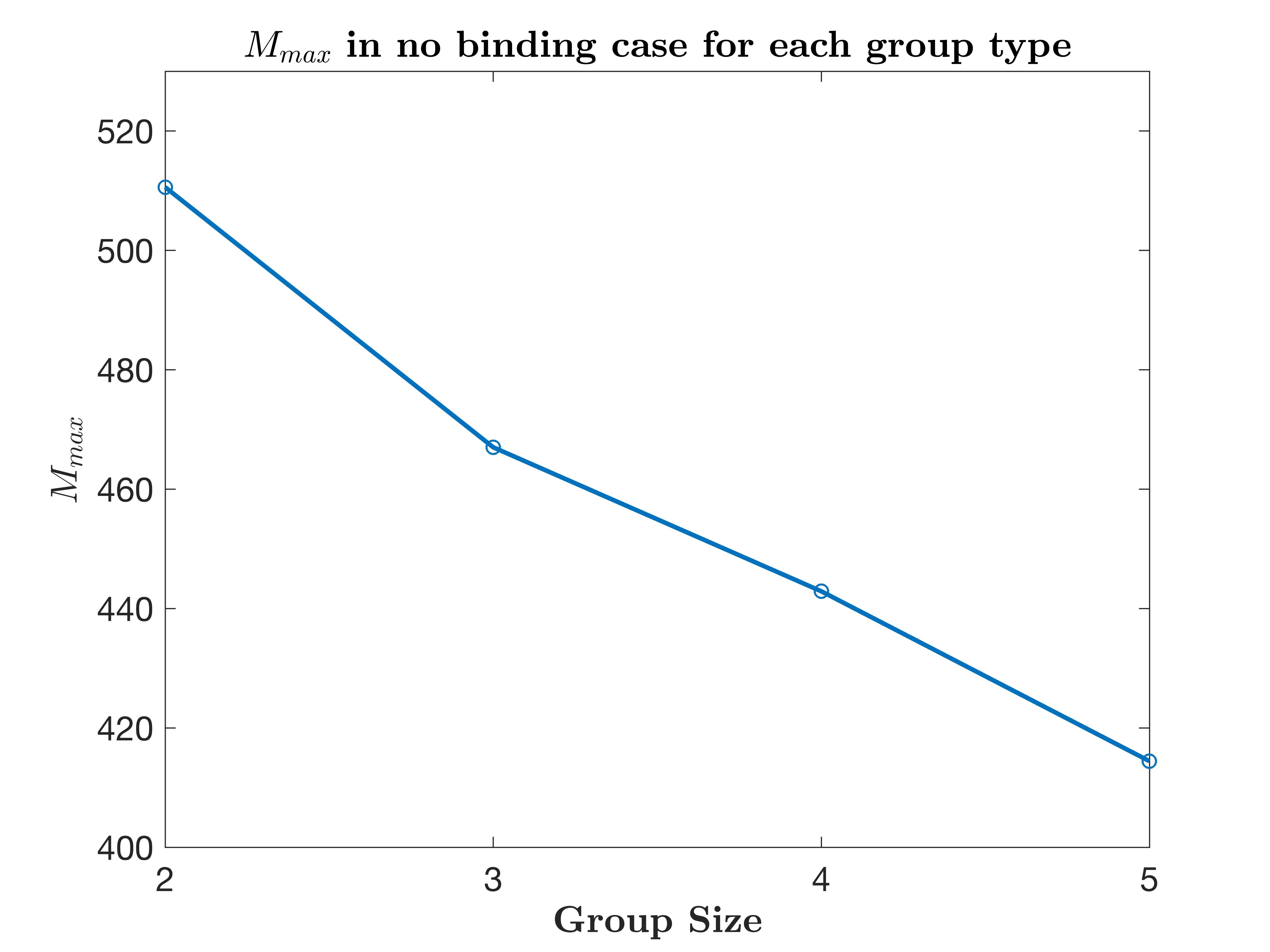}
    \subcaption{}
\end{subfigure}\quad
\begin{subfigure}{0.23\textwidth}\quad
    \includegraphics[width=0.9\linewidth]{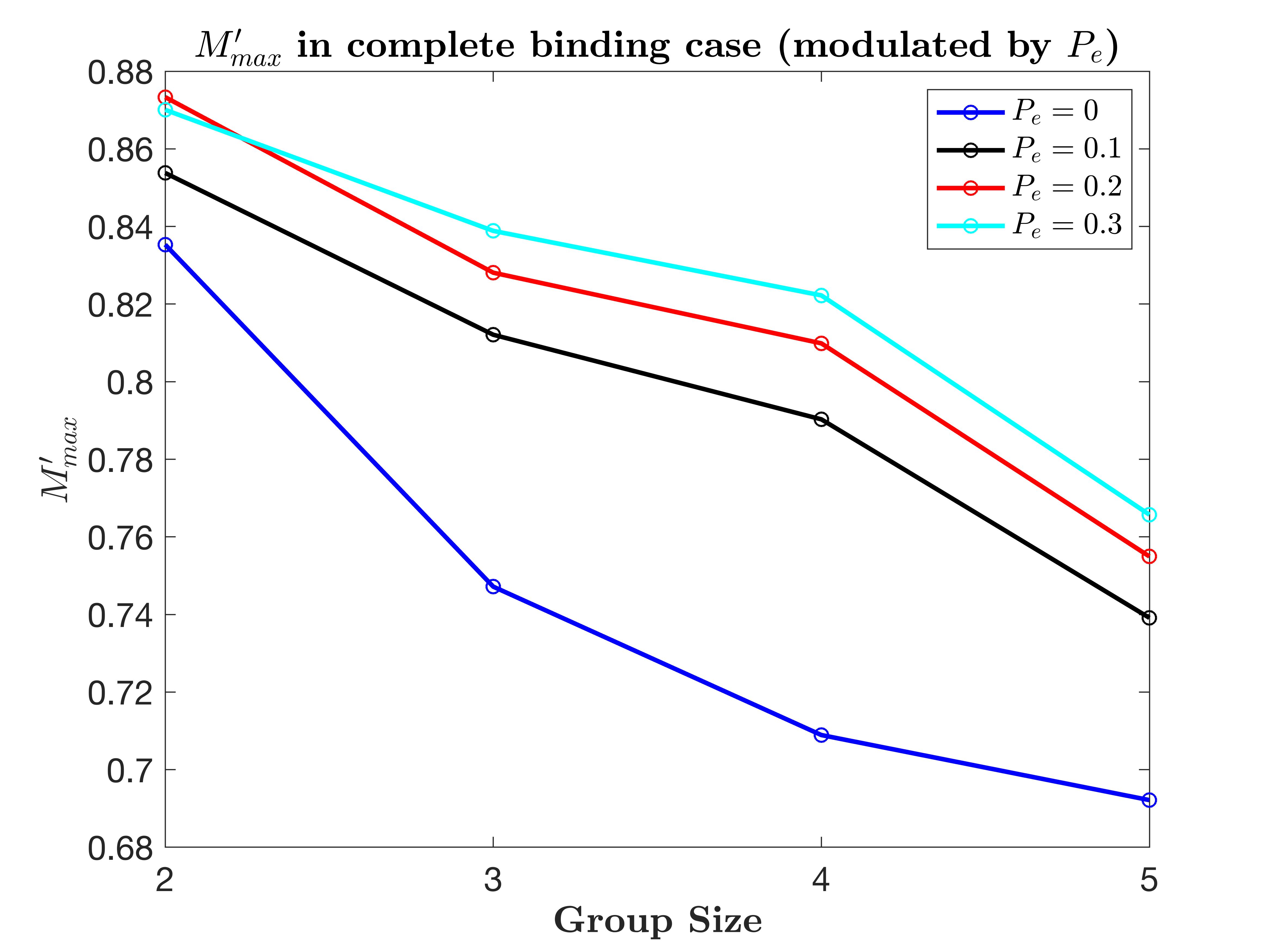}
    \subcaption{}
\end{subfigure} 
\begin{subfigure}{0.23\textwidth}
    \includegraphics[width=0.9\linewidth]{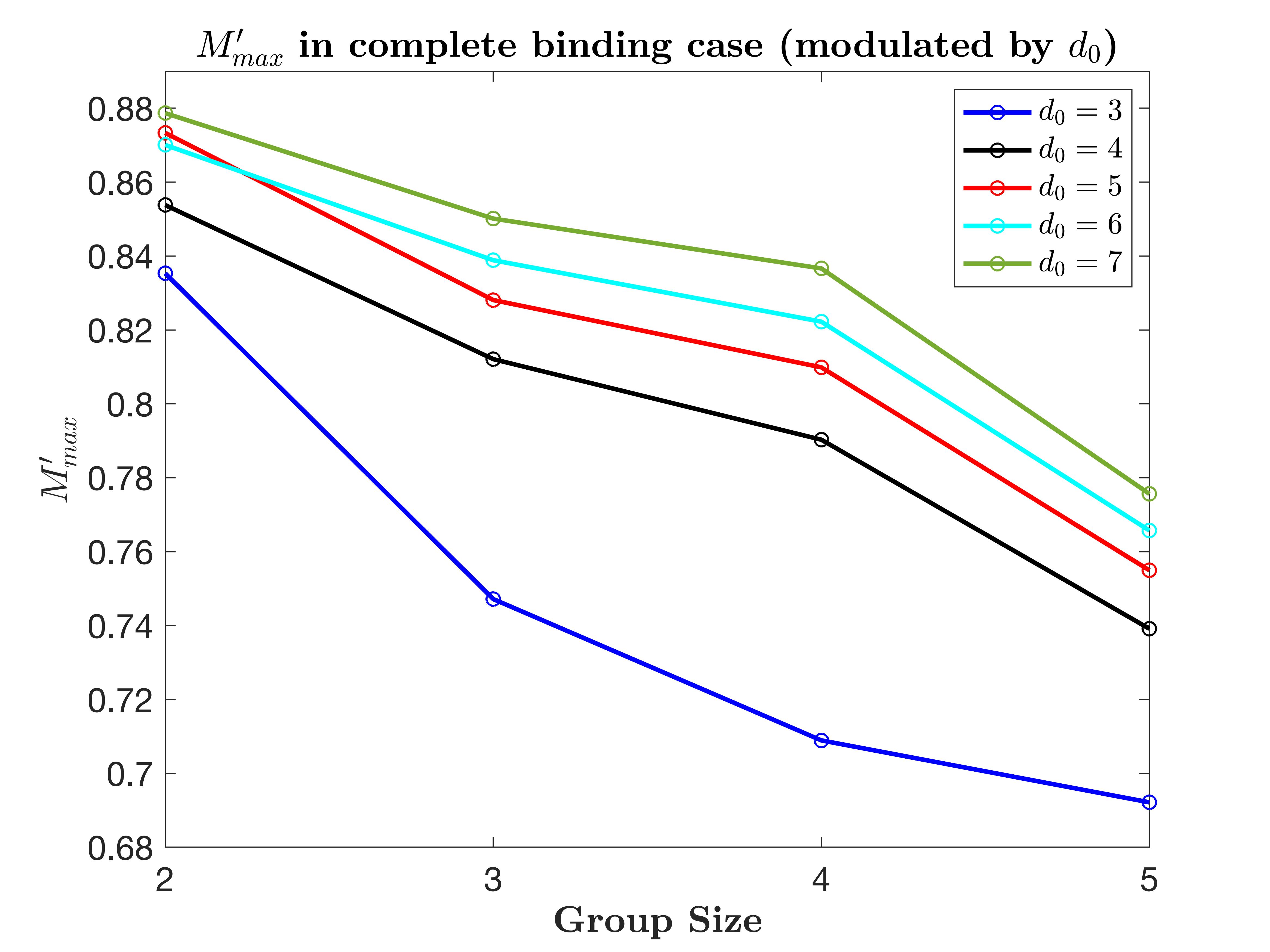}
    \subcaption{}
\end{subfigure}%
\caption{\label{fig:epsart}Time evolution of $M$ for the simulation of 2-agent group evacuation in the complete binding case is shown in the top panel.  The maximum of the system mixing index $M_{max}$ versus group size in both complete binding and no binding cases are shown in the middle panel. The relation between the normalized mixing index $M_{max}'$ and group size in the complete binding case with the binding strength modulated by both $P_e$ and  $d_0$ are in the bottom panel.}
\end{figure}

The dependence of $M_{max}$ on the group size under the complete binding is shown In Fig.12(b).  What we can see is that the mixing of different group members  is weakened to a greater extent as the group size increases.  The essential reason for this observation lies in the fact that the larger the size of each group, the less number of groups there will be, if the total number of agents is fixed. To get a more intuitional picture, we can analyze the spatial distribution of local mixing index at the moment when $M=M_{max}$. As shown Fig. 13, there are more grids of larger local mixing index with smaller group sizes. Indeed, it is less likely for agents from other groups to “invade” the area congregated by a larger group, under the setting of complete binding which drives agents of the same group to cluster together. 

\begin{figure}[htbp]

\begin{subfigure}{0.23\textwidth}
    \includegraphics[width=0.9\linewidth]{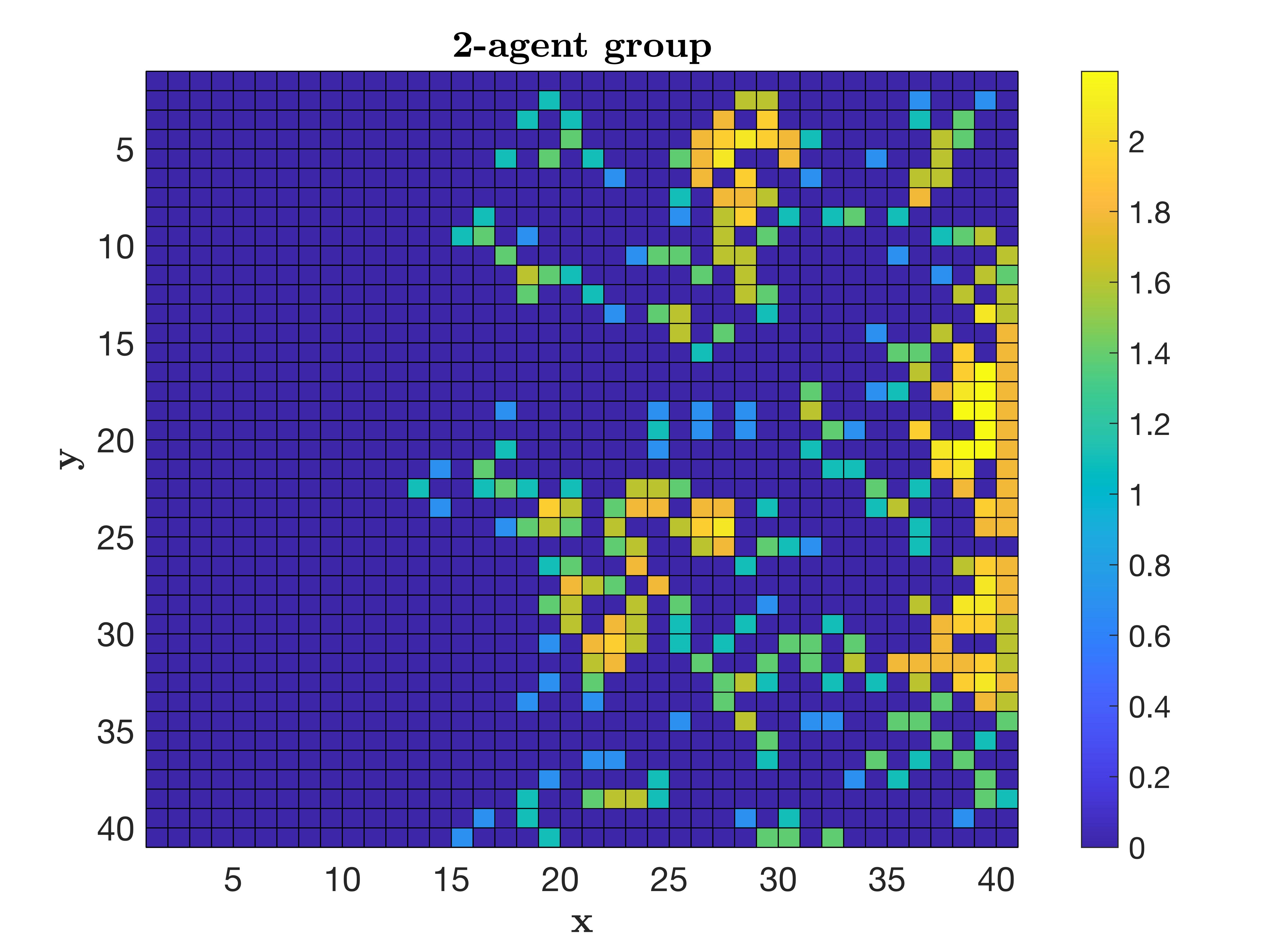}
    \subcaption{}
\end{subfigure} 
\begin{subfigure}{0.23\textwidth}
    \includegraphics[width=0.9\linewidth]{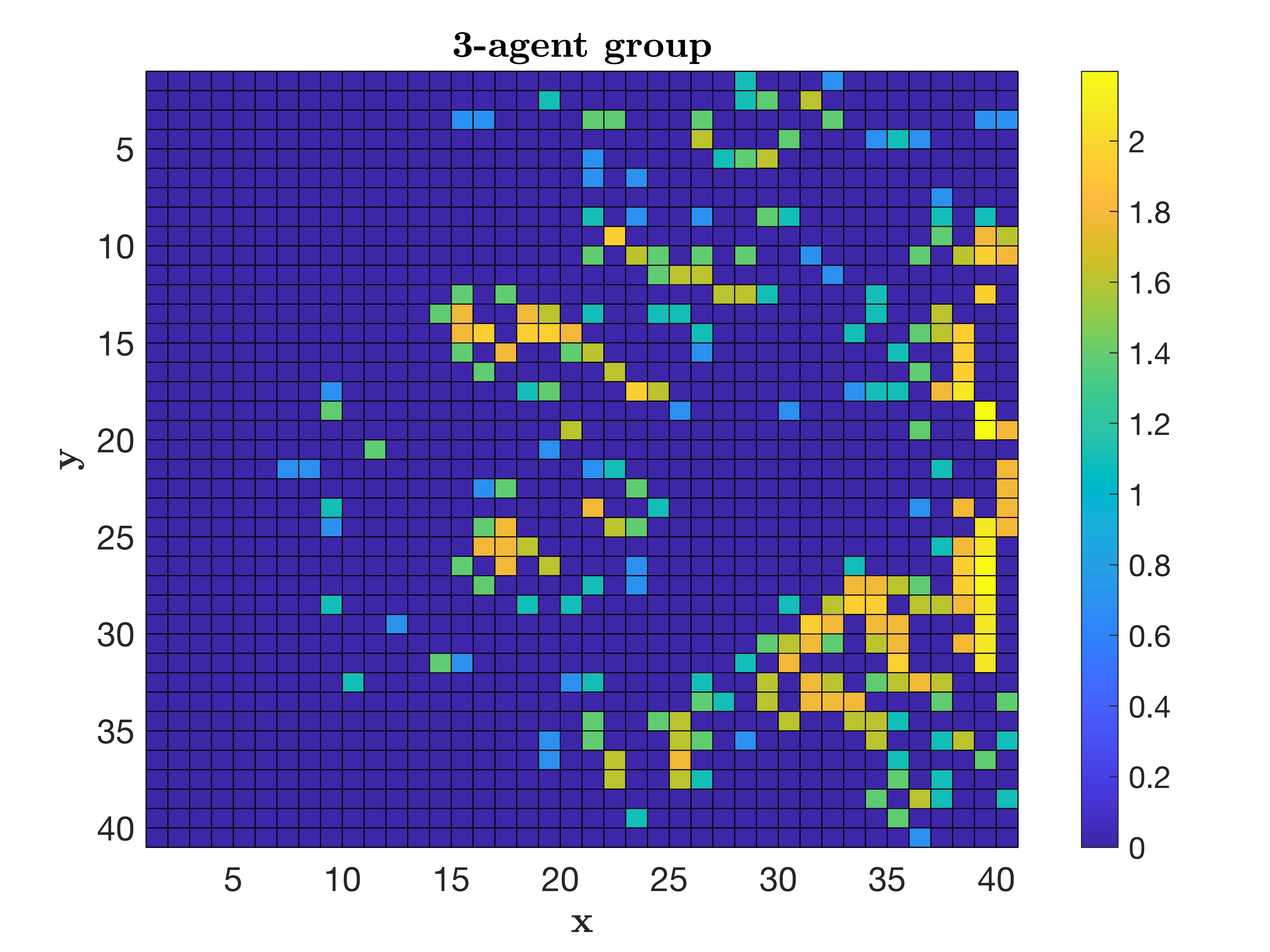}
    \subcaption{}
\end{subfigure} \quad
\begin{subfigure}{0.23\textwidth}
    \includegraphics[width=0.9\linewidth]{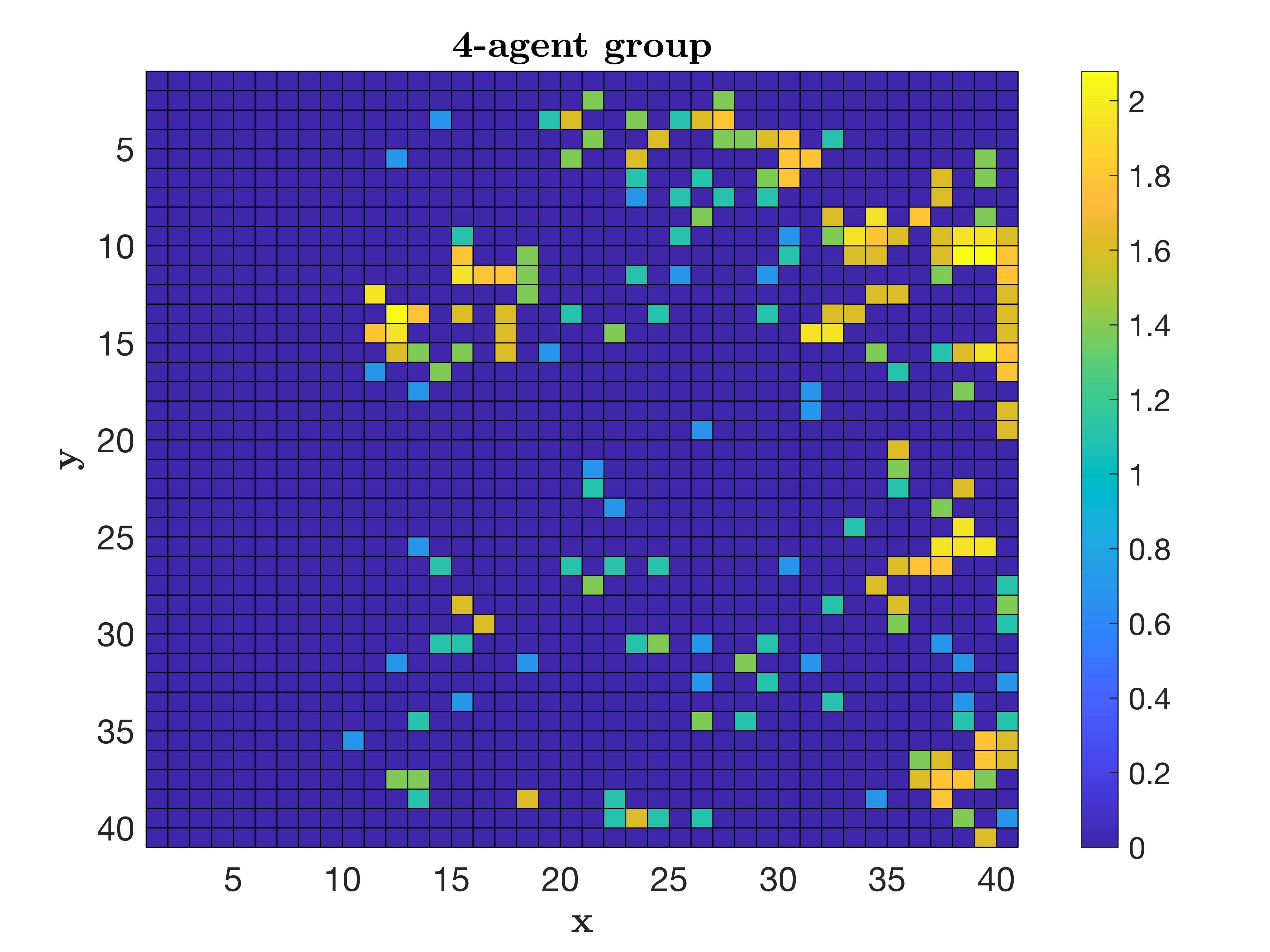}
    \subcaption{}
\end{subfigure}%
\begin{subfigure}{0.23\textwidth}
    \includegraphics[width=0.9\linewidth]{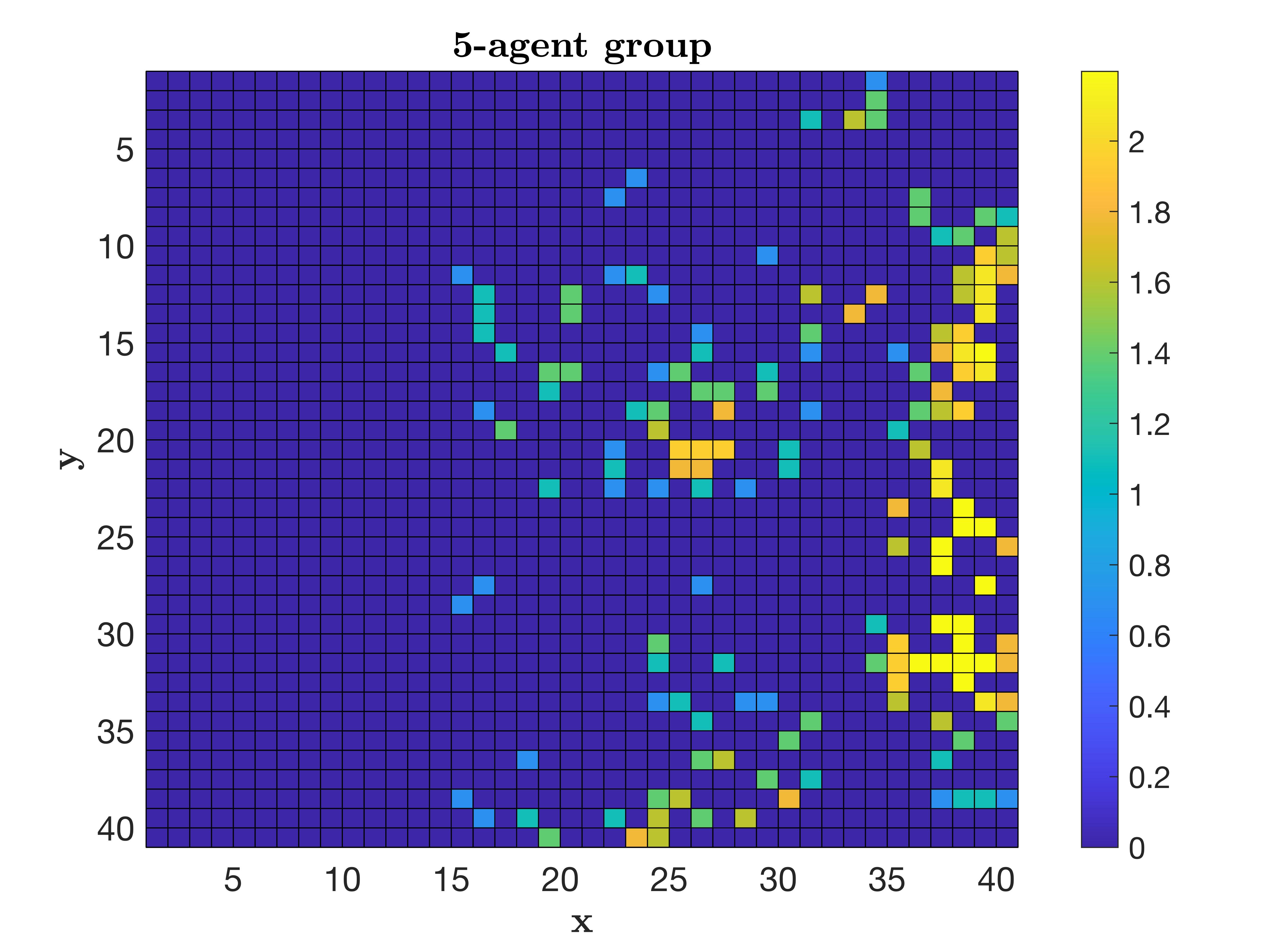}
    \subcaption{}
\end{subfigure} \quad
\caption{\label{fig:epsart}Spatial distribution of the local mixing index for all types of groups at the moment  when the system mixing index reaches its maximum.}
\end{figure}

Knowing that $M$ decreases as the group size increases, the answer to the negative dependence of evacuation time on group size lies in the relationship between the evacuation time and the system mixing index. When groups are highly mixed, it would be harder for followers to follow the leader since agents from other groups are more likely to be standing in their way. Then leaders will wait longer since group members are more likely to be left far behind. Yet mixing without the inclusion of complete binding cannot elongate the evacuation time. Evidence for this argument can be presented if we remove the binding completely. In this situation, the correlation between evacuation time and group size would be gone even though the inherent negative relation between $M$ and group size still exists, see Fig.9 and Fig.12(c).  Consequently,  we may conclude that the evacuation time is positively correlated with the mixing index only if the influence of complete binding exists.
Additionally, the negative dependence of the mixing index on group size strengthens as the strength of complete binding is increased. This argument may be demonstrated by investing the relation between the normalized system mixing index ($M'=M/M_{nobinding}$) with the group size, see Figs.12(e) and (f). Note that we can change the strength of complete binding either by tuning the critical distance $d_0$ or by changing the probability of malfunction of the waiting mechanism.

\subsection{Evacuation rate and the mixing index}
Having exploited the mixing index in the previous section, we try to find what more we can reveal the dynamics of evacuation with or without group structures from the mixing index. We define the rate of evacuation $R$ as the decrease in the number of agents in the evacuation area at a given time step. Intuitively we attempt to apply a constant scaling transformation to $M$ and to check if there would be a similarity between the time evolution pattern of the scaled M and that of $R$. The constant scaling factors are found through trial-and-error to be in the range of $0.01 - 0.015$ for the evacuation of individuals and all kinds of groups. As shown in Fig.14, we find that the evolution patterns of scaled $M$ and $R$ are similar regardless some large fluctuations in $R$. It suggests that the mixing index should contain useful information about the evacuation rate, which supports our optimistic envision about $M$, that it may serve as a powerful indicator for the dynamics of the evacuation system.

\begin{figure}[htbp]

\begin{subfigure}{0.23\textwidth}
    \includegraphics[width=0.9\linewidth]{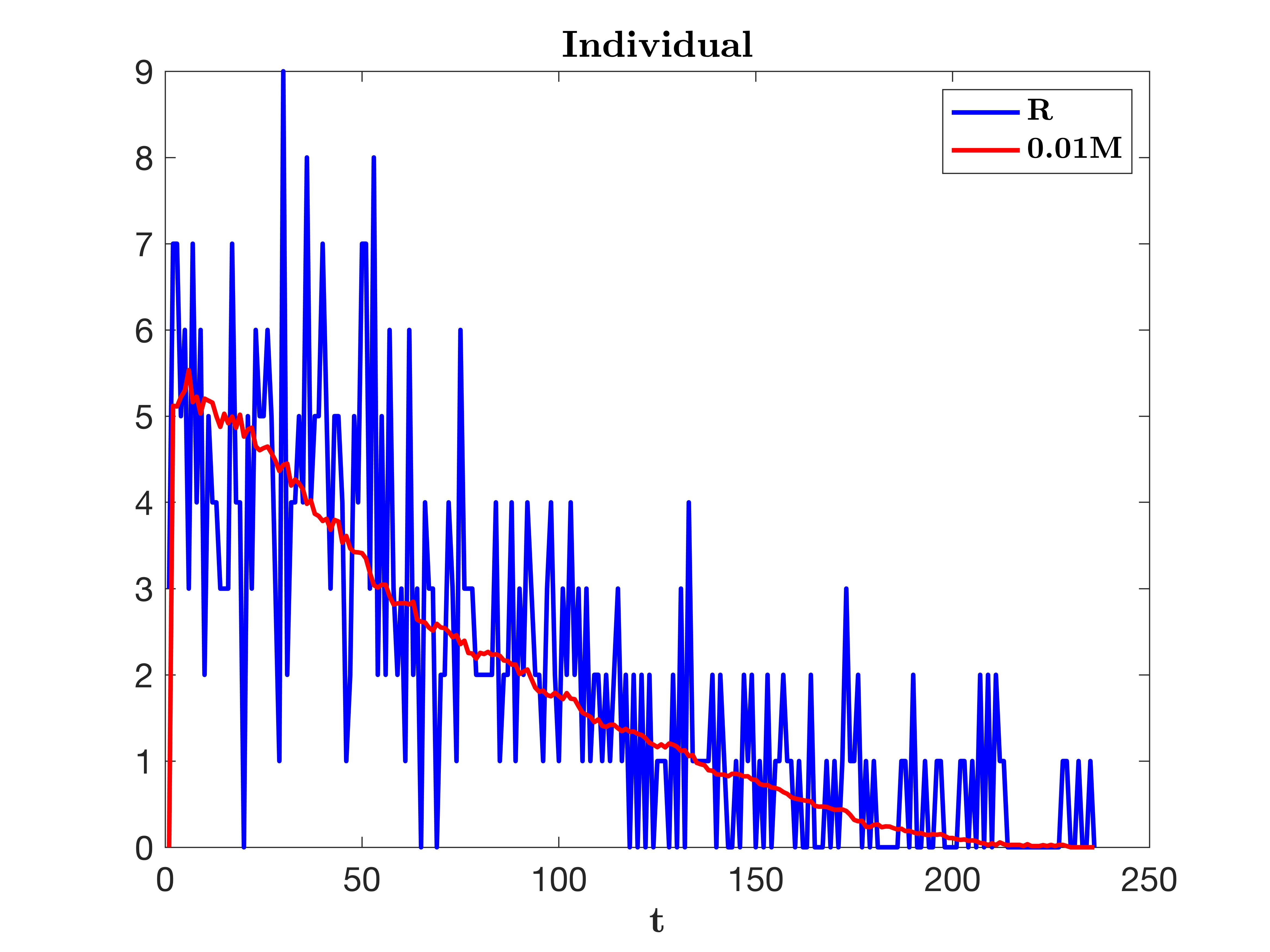}
    \subcaption{}
\end{subfigure} 
\begin{subfigure}{0.23\textwidth}
    \includegraphics[width=0.9\linewidth]{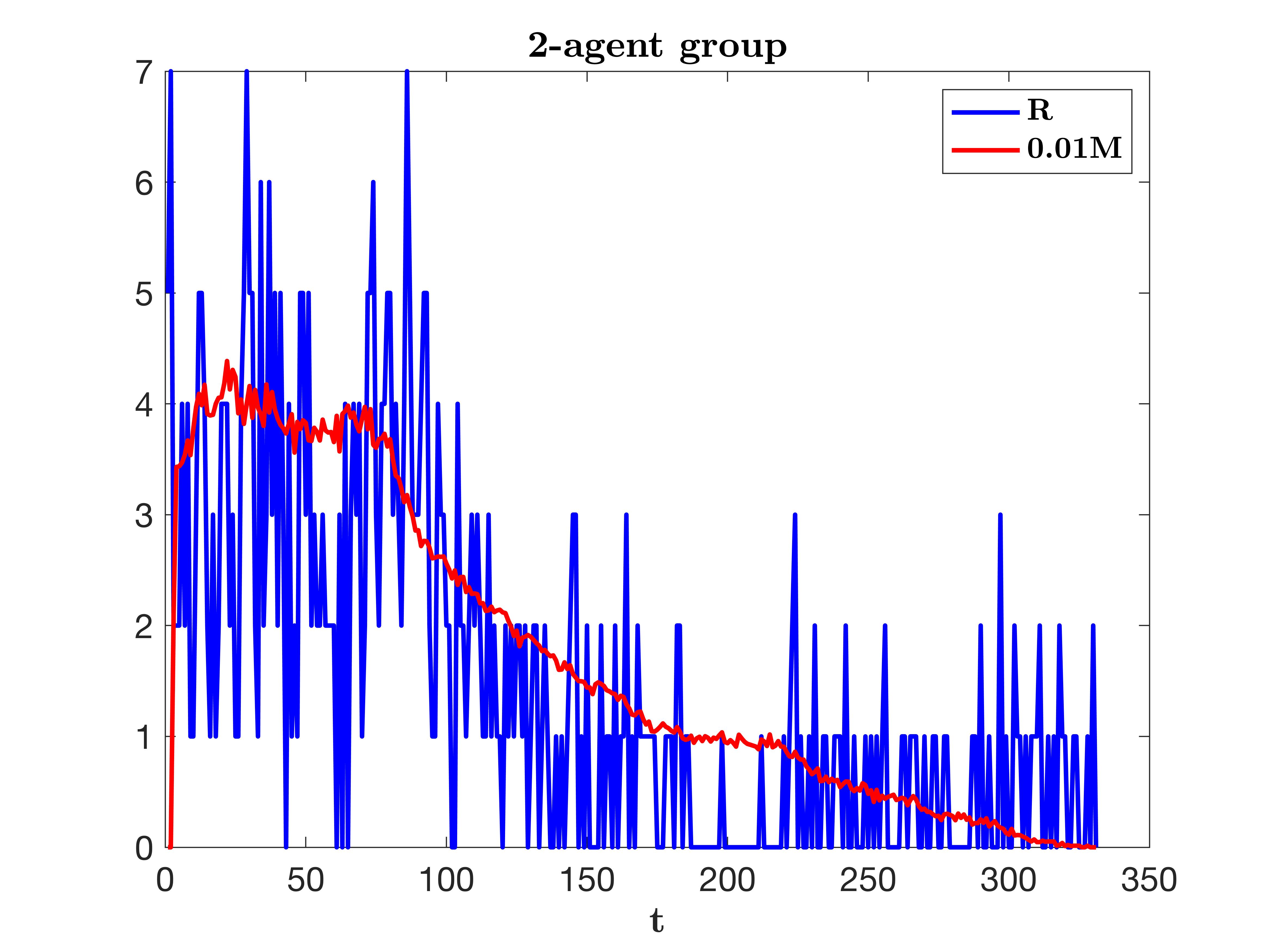}
    \subcaption{}
\end{subfigure} \quad
\begin{subfigure}{0.23\textwidth}
    \includegraphics[width=0.9\linewidth]{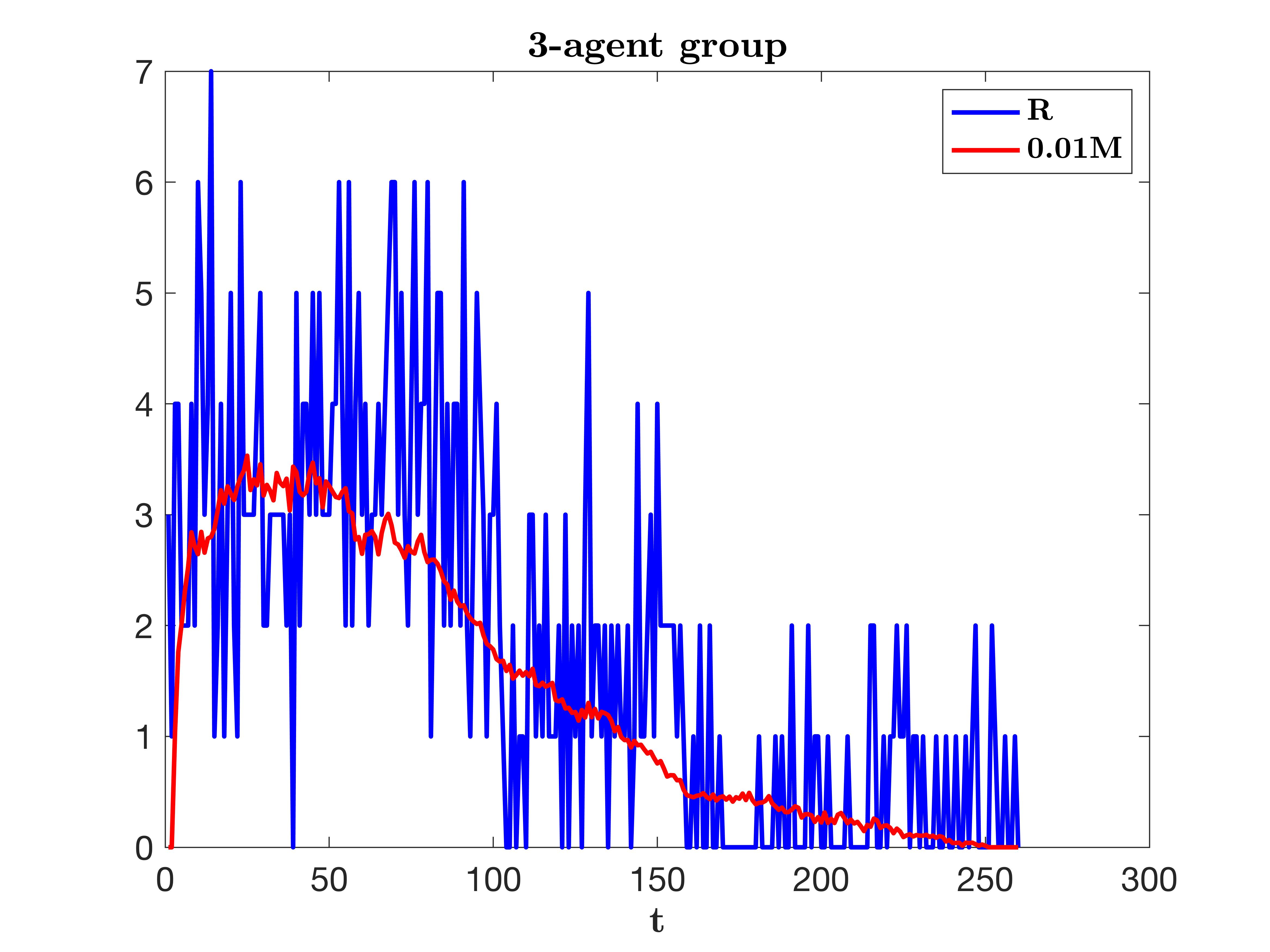}
    \subcaption{}
\end{subfigure}%
\begin{subfigure}{0.23\textwidth}
    \includegraphics[width=0.9\linewidth]{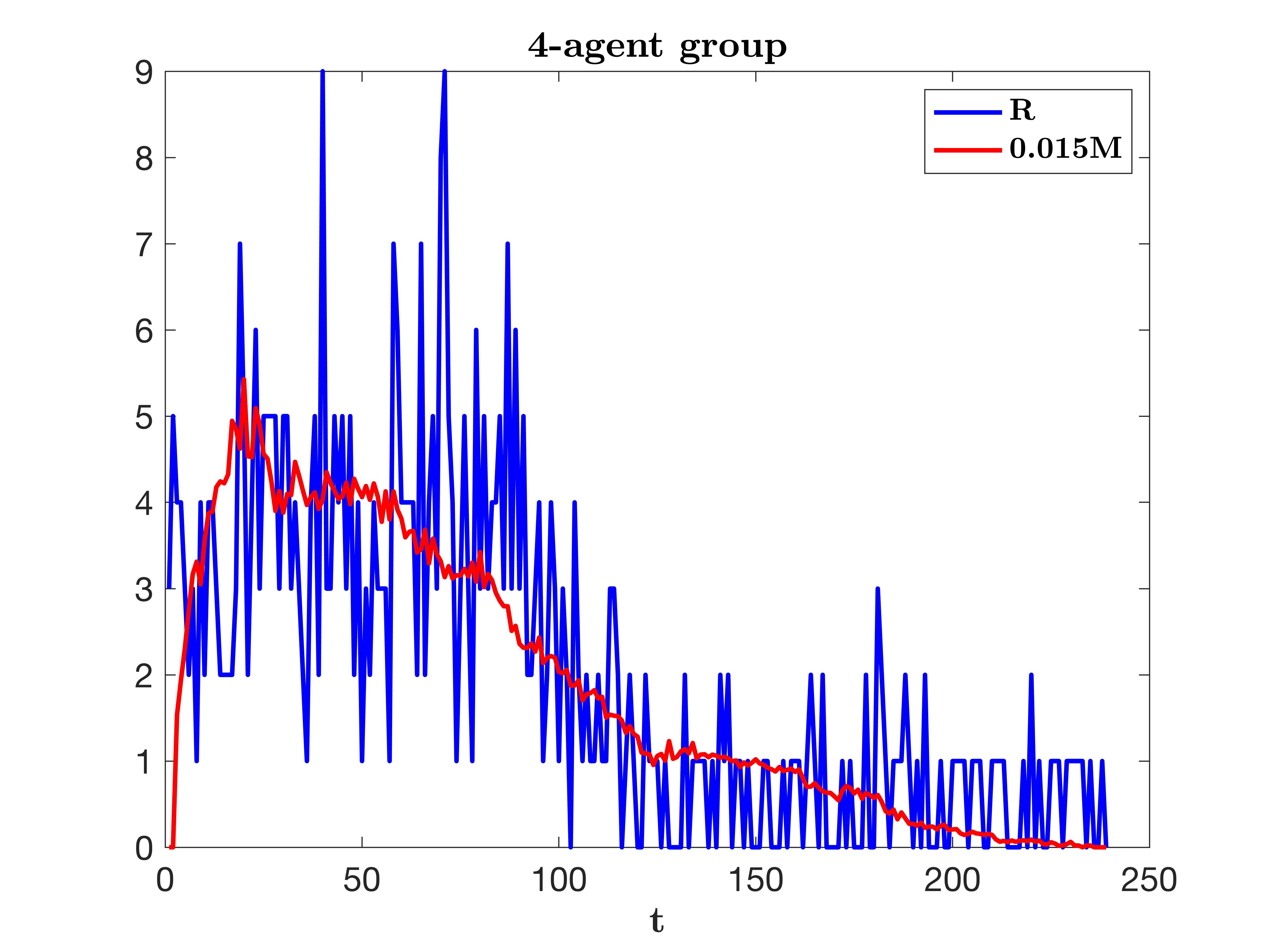}
    \subcaption{}
\end{subfigure} \quad
\begin{subfigure}{0.23\textwidth}
    \includegraphics[width=0.9\linewidth]{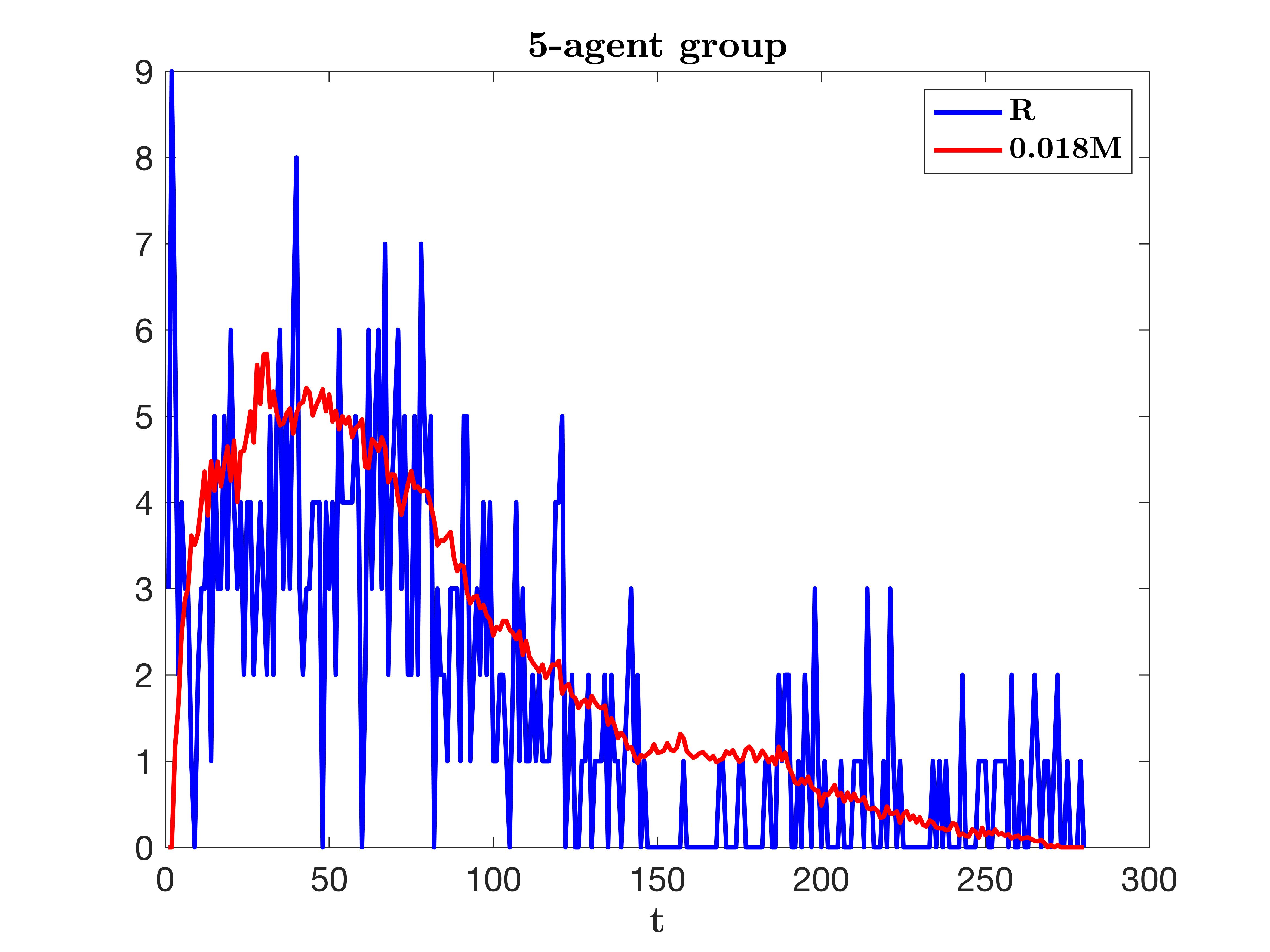}
    \subcaption{}
\end{subfigure}
\caption{\label{fig:epsart}Comparison of R-t (blue)  pattern with M-t (red) pattern ( being the constant scaling factor) for the evacuation of individuals and all types of groups.}
\end{figure}

\section{Conclusion}
In this paper, an extended floor-field cellular automaton with group structure, multi-speed-agents and local density-dependent dynamic field is constructed for the study of binding effects on evacuation systems. In particular, group structure is implemented by distinguishing between leaders and followers and by setting up a mutually binding mechanism. Our model is validated by calibration to the acknowledged results[14, 18]. 
Simulations show that when complete binding is taken into account, a negative dependence of evacuation time on the group size emerges. Whereas if the binding is incomplete or removed, there is no such dependence. To quantitatively analyze the cause for such a negative correlation, we proposed an entropy-like quantity, the local and system mixing index. The decrease of mixing index as the group size increases can be explained with a relatively more clear picture for the evacuation dynamics. In the meantime, we find that a higher level of mixing can give rise to a longer evacuation time if the complete binding works. Based on these two arguments,  we are able to clarify the reason behind the negative dependence of  evacuation time on the group size. 
Moreover, with a constant scaling transformation we find the similarity between the evolution pattern of evacuation rate and that of the mixing index, which suggests this entropy-like quantity can be an informative indicator for the evacuation system.
By now our conclusions are only theoretically self-consistent. Experiments employing real grouped personals with binding mechanisms specified are necessary before our theory can be fully verified and actually be applied to practical situations. Also, we think a thorough study about the mixing index $M$ as an indicator of the evacuation system is of great significance considering that it is a concept rooted in statistical physics and can be used to explain complex phenomena occurring in social systems.

\begin{acknowledgments}
We wish to acknowledge the support of the Graduate School of Frontier Science at the University of Tokyo for holding the UTSIP Program, and the support of the Department of Physics at Fudan University. 
\end{acknowledgments}

\bibliographystyle{unsrt}
\bibliography{Reference.bib}

\end{document}